\RequirePackage[hyphens]{url}
\documentclass[%
 reprint,
 amsmath,amssymb,
 aps,
]{revtex4-2}

\usepackage{graphicx,amssymb,amsmath,hyperref}
\usepackage{xfrac}
\usepackage{courier}
\usepackage{booktabs}
\usepackage[hyphens]{url}
\usepackage{comment}
\usepackage{lineno}
\usepackage{cancel}
\usepackage[normalem]{ulem}
\usepackage{xcolor}

\begin{document}
 
\title{Universal roughness and\\ the dynamics of urban expansion}


\author{Ulysse Marquis$^{1,2}$}

\author{Oriol Artime$^{3,4}$}

\author{Riccardo Gallotti$^1$}

\author{Marc Barthelemy$^{5,6}$}
\email{marc.barthelemy@ipht.fr}

\affiliation{$^1$ Fondazione Bruno Kessler, Via Sommarive 18, 38123 Povo (TN), Italy}
\affiliation{$^2$ Department of Mathematics, University of Trento, Via Sommarive 14, 38123 Povo (TN), Italy}
\affiliation{$^3$ Departament de Física de la Matèria Condensada, Universitat de Barcelona, 08028 Barcelona, Spain}
\affiliation{$^4$ Universitat de Barcelona Institute of Complex Systems (UBICS), Universitat de Barcelona, 08028 Barcelona, Spain}
\affiliation{$^5$ Universit\'e Paris-Saclay, CNRS, CEA, Institut de Physique Th\'eorique, 91191, Gif-sur-Yvette, France}
\affiliation{$^6$ Centre d’Analyse et de Math\'ematique Sociales (CNRS/EHESS) 54 Avenue de Raspail, 75006 Paris, France}

\begin{abstract}

Urban sprawl reshapes cities, yet its quantitative laws remain elusive. Analyzing built-up expansion in 19 cities (1985–2015) with tools from surface growth physics in radial geometry, we reveal anisotropic, branch-like growth and a piecewise linear scaling between area and population. We uncover a robust local roughness exponent $\alpha_{\text{loc}}\approx 0.54$, coexisting with variable $\beta$ and $z$. This unusual coexistence of universal and variable exponents offers a rare empirical testbed for nonequilibrium growth and an empirical basis for modeling urban sprawl.


\end{abstract}

\maketitle



Urban sprawl, the expansion of cities into surrounding regions, has been extensively studied from demographic and economic perspectives, primarily focusing on population growth \cite{gibrat1931, gabaix1999, verbavatz2020}, migration patterns, and socio-economic drivers \cite{bhatta2010,seto2011,batty2008}. Beyond population and economic drivers, urban expansion has led to environmental and sustainability challenges. The encroachment of urban areas onto agricultural land threatens food security, particularly in rapidly developing regions. Additionally, urban expansion near coastal zones increases the exposure of large populations to climate-related risks such as sea level rise and storm surges \cite{seto2011}.

High-resolution satellite data reveal a sharp acceleration in global urban expansion: urban land cover grew by $\sim 80\%$ between 1985 and 2015, at a rate of 9,687 km$^2$/year \cite{liu2020}, far exceeding earlier estimates \cite{seto2011}. Much of this growth has occurred in China, India, and Africa, driven by economic development and policy shifts, with GDP growth alone explaining nearly half the expansion in China \cite{seto2011}.

Despite its fundamental importance, the spatial dynamics of built-environment expansion remain poorly quantified. Most analytical approaches to urban sprawl lack empirical grounding \cite{ishikawa, bracken1992, simini2015, friesen, raimbault}, with the notable exception of \cite{capel-timms}, which analyzed long-term patterns in London and Sydney. Many studies focus on descriptive land cover and population trends rather than quantitative models \cite{seto2011}, and existing frameworks often prioritize economic or social factors over spatial dynamics \cite{bettencourt2013, fujita}.

A more systematic approach is needed to understand how cities grow \cite{batty2008}, to identify whether universal principles govern urban expansion, and—crucially—to determine the form of the evolution equation that describes urban sprawl, which remains an open question. Here, we introduce a new framework inspired by surface growth physics to study the spatiotemporal evolution of built-up areas. This framework, well established in systems such as turbulent liquid crystals \cite{takeuchi2010, Takeuchi_2012}, bacterial colonies \cite{santalla2018} and tumor growth \cite{bru1998, huergo2012}, and characterized by scaling laws and universality classes \cite{barabasi1995, odor2004}, has not yet been systematically applied to urban sprawl.

{\it Key issues---}Defining cities remains a key challenge \cite{dong2024}. Traditional definitions based on central flows \cite{fuas} are subjective and limit quantitative analyses of spatial growth. A major advance came with the City Clustering Algorithm (CCA) \cite{cca}, which uses percolation theory to identify urban areas as clusters of built-up space. Wilson et al. \cite{wilson2003} distinguished three urban growth types: infill (within existing fabric), expansion (at the fringe), and outlying growth, including isolated, linear, and clustered branch growth. These reflect interacting mechanisms of sprawl. We use the CCA to extract the largest connected component—the giant cluster \cite{percolation}—as a robust object to study urban growth.

A second key issue is the choice of an appropriate time parameter. Population $P$ naturally serves as a clock for tracking urban evolution \cite{bhatta2010,batty2008}. Unlike chronological time, it helps mitigate the effects of external disruptions—such as wars, epidemics, or short-term economic fluctuations—on growth dynamics. A quantitative analysis of urban sprawl should therefore examine the evolution of spatial variables (e.g., built-up area, average radius) as functions of $P$.

A third challenge is data availability. While population records span multiple decades, built-up area data is often incomplete or limited in coverage. Since our study requires both temporal built-up area and historical population data for urban areas, the data selection process is highly constrained. Additionally, we prioritized cities with relatively isotropic growth and minimal geographical constraints to enable a radial surface analysis. Despite these strong constraints, we successfully compiled a dataset covering most continents. We identified 19 major cities worldwide that meet these criteria, forming the basis of our analysis. Built-up area data was sourced from the WSF Evolution dataset \cite{wsfevo}, while population data was obtained from the World Population Prospects \cite{worldpop} (see SM for details).

\begin{figure*}[htp!]
    \centering
    \includegraphics[width=1\linewidth]{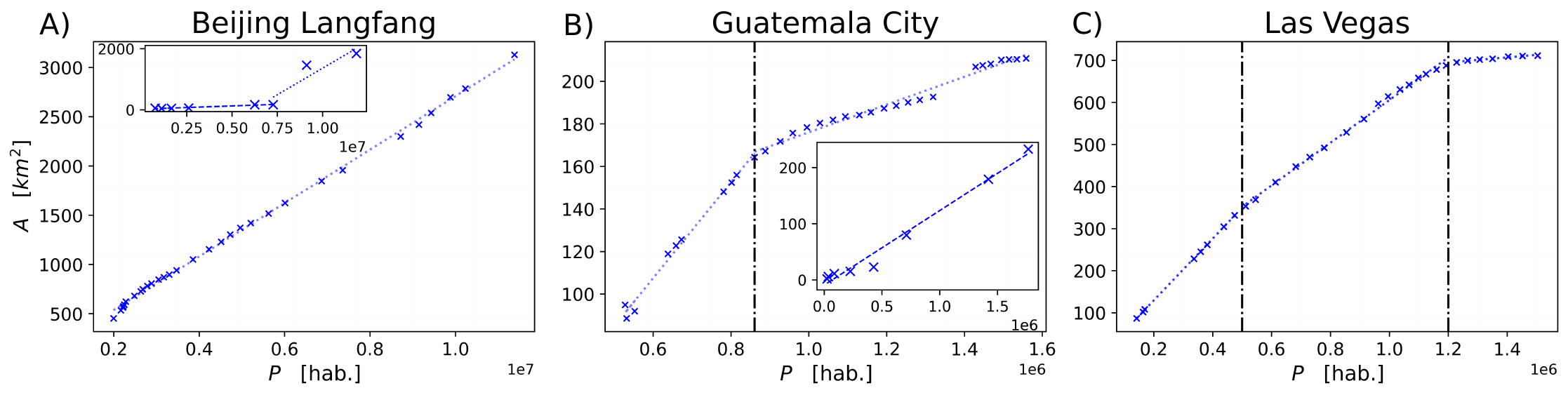}
    \caption{\small{
    Different growth patterns of the largest connected component (LCC) area vs. population (1985–2015): (A) linear scaling at constant density (Beijing, $\sim$3,685 inh./km$^2$); (B) piecewise linear growth with increasing density (Guatemala City, from 4,437 to 15,314 inh./km$^2$); (C) saturation plateau (Las Vegas, up to 16,782 inh./km$^2$). Inset: Historical trends (1800–2000). Beijing shows a density drop from 48,393 to 2,882 inh./km$^2$ around 7M population, while Guatemala City maintains near-constant density (7,576 inh./km$^2$).}
}
    \label{fig:area_vs_pop_linear}
\end{figure*}

 {\it Area growth speed and anisotropy---} We begin by examining how total built-up area scales with population. Historical data from 1800 to 2000 \cite{atlasurbanexpansion} show that early urban areas largely coincided with the largest connected component (LCC). We identify piecewise linear trends with breakpoints indicating shifts in average population densities (see insets in Fig.~\ref{fig:area_vs_pop_linear} and Fig.~S1 in the Supp. Mat. \cite{Supplemental}). In London, density dropped from 18,000 to 1,800 inhabitants/km$^2$ around 1914; in Paris, a similar decline occurred between 1979 and 1987, from 8,500 to 810 inhabitants/km$^2$. These findings contrast with cross-sectional scaling relations from \cite{bettencourt2013} (see Fig.~S2 in the SM). From 1985 to 2015, we observe similar patterns across 19 cities (see Figs.~S4--S5 in the SM), motivating the approximation $P \approx A \cdot P_{\text{tot}} / A_{\text{tot}}$, where $P_{\text{tot}}$ is total population of the whole urban area $A_{\text{tot}}$, and $P$ is the population of the LCC (and $A$ its area). The relationship between $A$ and population $P$ reveals three growth patterns (Fig.~\ref{fig:area_vs_pop_linear}): (A) linear scaling with constant density, as in Beijing ($\sim 3{,}685$ inhabitants/km$^2$); (B) piecewise linear with increasing density, as in Guatemala City; and (C) saturation, as in Las Vegas, where growth halts due to topographical or policy constraints \cite{lasvegas2,lasvegas3}.

Assuming isotropic growth with uniform density, $A \sim P$ implies $r(P) \sim P^{1/2}$, consistent with average trends \cite{lemoy2021radial}. To quantify anisotropy, we define $r(\theta, P)$ as the radius in a sector $[\theta, \theta + \delta \theta]$ centered on the city (see SM). We find $r(\theta,P) \sim P^{\mu(\theta)}$, with $\mu(\theta) \in [0, 2.5]$ (see Fig.~S6 in the SM). Large exponents correspond to rapid growth along specific directions, often roads (see videos \cite{videos}). Deviations from the isotropic case $\mu = 1/2$ are used to quantify anisotropy (Fig.~S7 in the SM). Six cities exhibit low relative dispersion (e.g., Changzhou), while others such as Paris are highly anisotropic (see Figs.~S8 and S9 in the SM for maps of Changzhou and Kolkata) \cite{percolation}.


{\it Growth mechanisms---}Our dataset reveals two distinct modes of surface growth: (i) \textit{local growth}, occurring at the urban interface in a manner similar to diffusion, and (ii) \textit{coalescence}, where expansion results from the merging of larger surrounding clusters (see growth videos~\cite{videos}), as conjectured in~\cite{herold2005}. To quantitatively distinguish these mechanisms, we write the new area that appears at time $t+1$, denoted $\delta A(t+1)$, as the sum of two terms:
\begin{align}
    \delta A(t+1) = C_o(t) + C_n(t),
\end{align}
where $C_o$ corresponds to clusters that already existed at time $t$ (but were not connected to the largest connected component, LCC), and $C_n$ denotes newly built areas that did not exist at time $t$.

\begin{figure}
    \centering
    \includegraphics[width=0.9\linewidth]{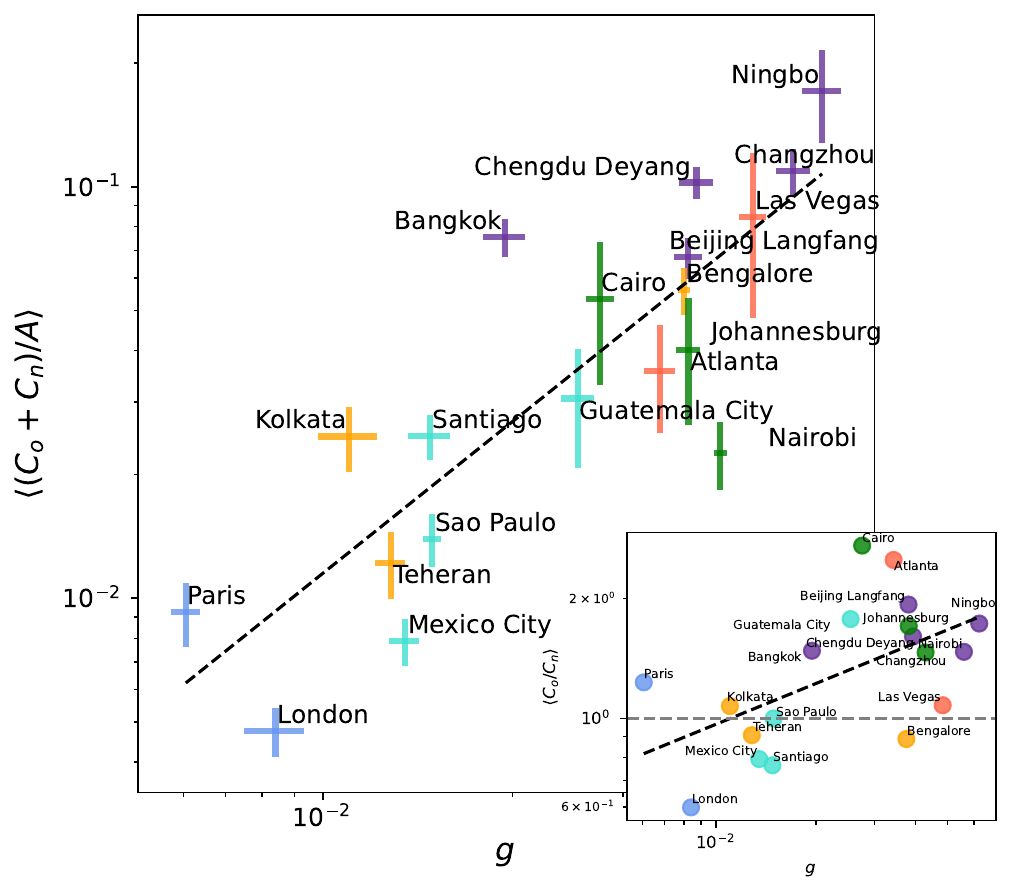}
    \caption{\small{Relative aggregated area as a function of demographic pressure, exhibiting a clear increasing trend. The dashed line represents a power-law fit with an exponent of approximately $1.22$ (with $R^2=0.71$), provided as a visual guide. While the population growth rate changes from $\approx 1\%$ for London to $6\%$ for Ningbo, the relative size of aggregates increases by more than one and a half order or magnitude (the error bars represent the standard error). Inset: average ratio of coalesced built sites ($C_o$) relative to newly built sites ($C_n$). An increasing trend quantified by a Spearman's correlation of~$0.51$ is observed. The dark dashed line represents a power-law of exponent~$0.33$. Color code : geographic region.}}
    \label{fig:clustersize_vs_pressure}
\end{figure}

To explore the relationship between growth patterns and urban expansion dynamics, we first plot (Fig.~\ref{fig:clustersize_vs_pressure}) the temporal averages (denoted by $\langle\cdot\rangle$) of both components in $\delta A(t+1) = C_o(t) + C_n(t)$, normalized by $A(t)$, against the demographic pressure, quantified by the average population growth rate $g = \langle dP_{\text{tot}} / (P_{\text{tot}}\, dt) \rangle$. A high population growth rate is expected to exert pressure that accelerates urban expansion, which we indeed observe here. In addition, the inset of Fig.~\ref{fig:clustersize_vs_pressure} shows the ratio $\langle C_o / C_n \rangle$ versus $g$, indicating that coalescence increases with demographic pressure and becomes predominant for $g > 10^{-2}$ (we show additional results in Fig.~S10 in the SM). 

This trend is indeed confirmed in Fig.~\ref{fig:clustersize_vs_pressure}, where we observe that $\langle \delta A/A\rangle$ increases with $g$ (a power-law fit suggests an exponent of approximately $1.22$). When demographic pressure is low, surface expansion primarily occurs through local agglomeration, with successive layer additions resembling surface growth processes in physics (as exemplified by Kolkata in Fig.~S9 of the SM). Conversely, under high demographic pressure, the LCC expands more abruptly by also assimilating large surrounding clusters—sometimes comparable in size to the LCC itself— exemplified by Ningbo, Las Vegas, and Cairo in Fig.~S10 of the SM. This mechanism closely parallels growth and coalescence patterns observed in animal dispersal \cite{shigesada1997,carra2017}.

{\it Roughness exponents---} In surface growth physics, it is standard practice to measure various exponents that characterize the evolution and roughness of interfaces, enabling the identification of distinct universality classes~\cite{barabasi1995,odor2004} (see Section V in \cite{Supplemental} for a brief introduction). A well-established approach involves studying the width $w(L,t)$ of a one-dimensional growing surface over time for a system of linear size $L$, and a key result is that it follows the Family-Vicsek scaling relation $w(L,t) \sim t^\beta F(L/t^{1/z})$, where the scaling function behaves as $F(u \ll 1) \sim u^\alpha$. Here, $\beta$ is the growth exponent, describing how quickly roughness develops, while $\alpha$ is the roughness exponent, characterizing the surface's spatial irregularity at macroscopic scale. The correlation length $\xi$ evolves as $\xi \sim t^{1/z}$, where $z$ is the dynamic exponent, indicating the rate at which roughness saturation is reached (when $\xi \sim L$, i.e at crossover time $t_\times \sim L^z$). Most of these studies have focused on the one-dimensional case with a band geometry which is not directly relevant here. Radial generalizations, more relevant to urban expansion, remain less explored but have been considered in \cite{takeuchi2010, Takeuchi_2012,  huergo2011, marcos2025, kapral1994,escudero2008,rodriguez2011,masoudi2012, ferreira2006,santalla2014, escudero2012, santalla2015, takeuchi2011, De_Nardis_2017}. 
In a radial geometry, where a surface expands on a two-dimensional plane from a central point, several issues arise. First, the system scale $L$ used in traditional band geometry systems is not well-defined, requiring alternative methods \cite{domenech2024}, such as power spectrum collapse \cite{huergo2011, ramasco2000} or height-height correlation function \cite{marcos2025}, in order to determine the global roughness exponent $\alpha$. Measuring growth via the total interface perimeter presents several challenges: first, obtaining a well-defined measure may require an ultraviolet cutoff for fractal boundaries \cite{santalla2024}; second, its time evolution must be properly accounted for. If the size of the system grows at a faster rate than the correlations, saturation never occurs \cite{santalla2014, domenech2024}. Second, anisotropy, often overlooked in previous studies, complicates growth characterization and make it difficult to extend correlation-based or spectral methods \cite{santalla2024,marcos2025,huergo2011}. 

In an isotropic system, the average radius, $\overline{r} = \frac{1}{2\pi} \sum_{\theta} r(\theta)$, effectively describes expansion. However, in the presence of anisotropy, this measure loses significance, complicating the task of measuring the growth exponent $\beta$ or correlation functions. In order to remedy to this issue, a solution consists in introducing $N$ sectors of size $\Delta\theta$, such that $N\Delta\theta=2\pi$. The average width is then defined as an average over all sectors
\begin{align} \label{eq:localwidth}
    w^2(\overline{\ell},P)=\frac{1}{N}\sum_{i=1}^N\langle[r(\theta,P)-\langle r\rangle_i]^2\rangle_i
\end{align}
where $\langle\cdot\rangle_i$ is the average computed over all angles in a given sector $i$ ($\langle r\rangle_i$ is then the average radius over the sector $i$). For a given sector of average radius $\langle r\rangle_i$, we thus have an average arc length $\overline{\ell}(\theta)=\langle r\rangle_i\Delta\theta$ introducing in the system a new length scale.

\begin{figure*}[htp]
    \centering
    \includegraphics[width=0.7\linewidth]{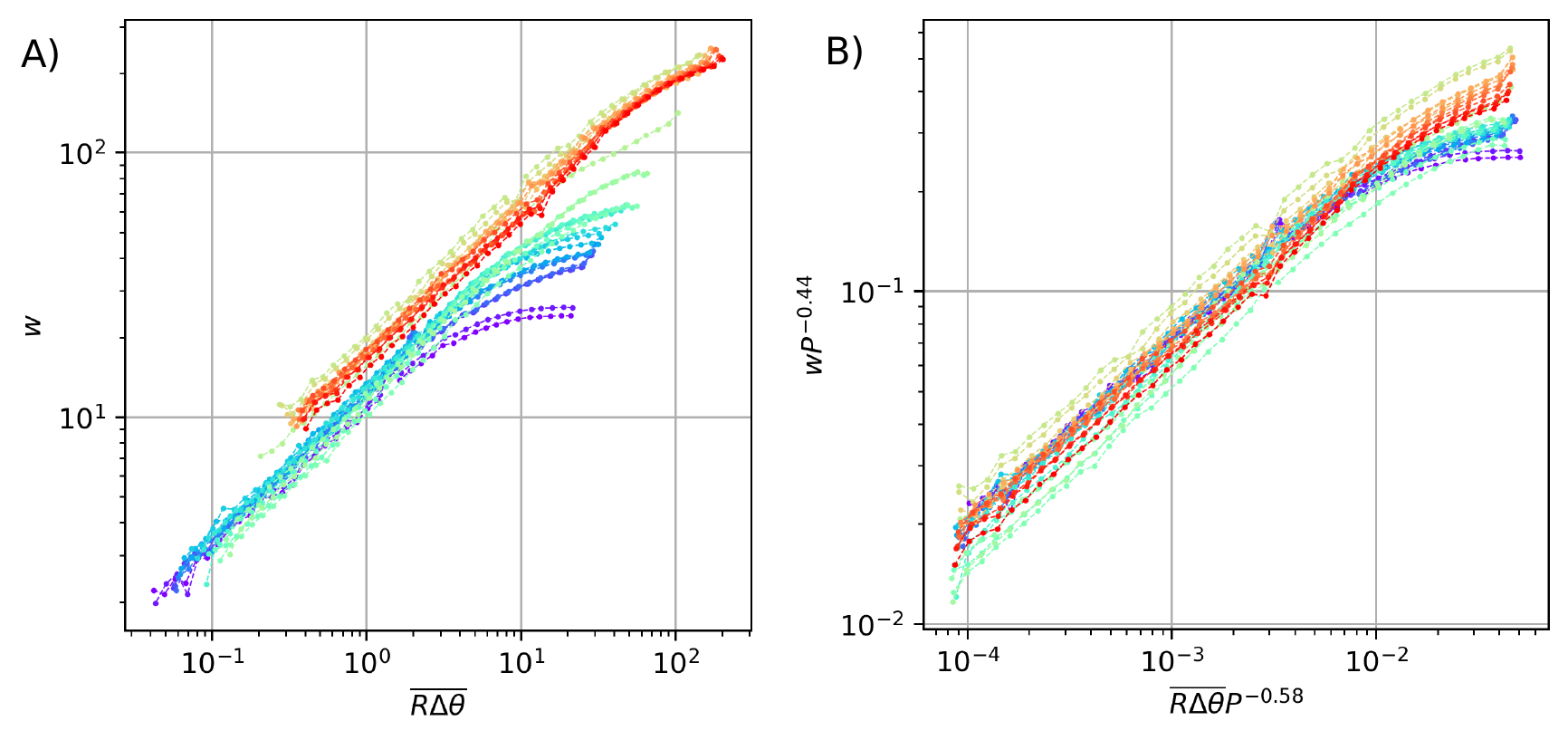}
    \caption{\small{(A) Interface width $w$, computed for angular sectors $\Delta \theta$, plotted against the average arc length $\overline{\ell}=R\Delta\theta$ (where $R=\langle r\rangle_i$ for each sector $i$) 
    for the city of Ningbo, China, for each year between 1985 (purple) and 2015 (red).  (B) According to the scaling form in Eq.~\ref{eq:scaling}, these curves should collapse onto a single master curve when plotting $wP^{-\beta}$ against $R\Delta\theta P^{-1/z}$. This collapse allows for the determination of the exponents $\beta$ and $z$.}}
    \label{fig:collapse}
\end{figure*}

In the isotropic case, $\langle r \rangle_i$ is independent of $i$, and the surface boundary is, on average, a circle. However, in the general anisotropic case, as seen in cities, the average interface deviates from a circular shape. We adopt the following anomalous scaling ansatz \cite{ramasco2000}, as described in studies on anisotropic radial growth \cite{domenech2024}
\begin{align}
    w(\overline{\ell},P)=P^\beta F\left(\frac{\overline{\ell}}{P^{1/z}}\right)
    \label{eq:scaling}
\end{align}
where $\overline{\ell}$ is the average arc length of a sector of aperture $\Delta \theta$ defined above, and where $F(u \ll  1) \sim u^{\alpha_{\text{loc}}} $
with $\alpha_{\text{loc}}$ being the {\it local} roughness exponent (see SM for details). In our case, we do not expect the scaling to hold outside the $u \ll 1$ regime: as $\Delta \theta$ increases, the angular-resolved width will begin to reflect both curvature and anisotropy.
The scaling form Eq.~\ref{eq:scaling} suggests that plotting $wP^{-\beta}$ against $\overline{\ell} P^{-1/z}$ should result in a data collapse onto a single curve (for small enough angular aperture $\Delta \theta$), allowing for the determination of the exponents $\beta$ and $z$ (see SM for more details). We illustrate this method on the case of Ningbo (China) in Fig.~\ref{fig:collapse}.  
Once the collapse is achieved, in the first regime where $\overline{\ell} P^{-1/z}\ll 1$, we expect a slope equal to $\alpha_{\text{loc}}$, following the scaling relation $w(\bar{\ell}) \sim \bar{\ell}^{\alpha_{\text{loc}}}$. A linear regression in log-log space allows us to determine the value of $\alpha_{\text{loc}}$. For the cities analyzed in this study, we find a consistent local roughness exponent of $\alpha_{\text{loc}} = 0.54 \pm 0.03$, indicating minimal variation across cities. This value differs from the global roughness exponent $\alpha$, estimated here via the relation $\alpha = \beta z$, which varies widely in the range $[0,2]$. In nearly all cases—except for cities with $\beta \approx 0$ (and for Kolkata)- we observe $\alpha > \alpha_{\text{loc}}$, with an average value of $\alpha = 0.776 \pm 0.475$ (see Table S1 in \cite{Supplemental} for details). This systematic difference indicates that the scaling is intrinsically anomalous: the standard Family–Vicsek scaling ansatz is not satisfied, and the interface does not fall into the super-rough regime~\cite{ramasco2000}.

The universality of this $\alpha_{\text{loc}}$ value suggests that the LCCs of cities share similar properties. Hence, the mechanisms governing local roughness appear fundamentally alike across all urban expansions studied here, despite the diversity of these cities and their growth dynamics. We note that the observed exponent close to~$1/2$ could suggest a purely probabilistic origin, consistent with arguments in~\cite{bernard2024,private}, although this point warrants further investigation to be fully elucidated (see Fig. S12 in the SM for more details and discussions on this point).

\begin{figure}[htp]
    \centering
    \includegraphics[width=0.7\linewidth]{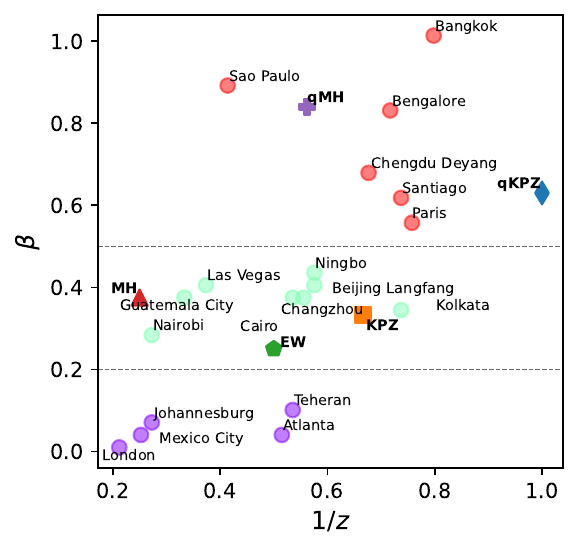}
    \caption{\small{Exponent $\beta$ versus $1/z$ for cities in our dataset. We can distinguish 3 broad groups according to the value of $\beta$ (shown in different colors): purple for $\beta$ small, green for $\beta\approx 0.37$, and red for $\beta>0.6$. We also indicate usual universality classes values for $\beta$ and $z$: Edwards-Wilkinson (EW) with $\beta=1/4$, $z=2$ (and $\alpha=1/2$), Kardar-Parisi-Zhang (KPZ) with $\beta=1/3$, $z=3/2$ (and $\alpha=1/2$), and Mullins-Herring (MH) with $\beta=3/8$, $z=4$ (and $\alpha=3/2$), along with two quenched universality classes at the depinning transition, qKPZ ($\beta=0.63$, $z=1$) and qMH ($\beta=0.84$, $z=1.78$).}}
    \label{fig:exponents}
\end{figure}

The value of $\alpha_{\text{loc}}$ alone does not allow to identify any universality class and we have to measure $\beta$ and $z$. In contrast, for these exponents our dataset reveals substantial variability in the exponents $\beta$ and $z$ across different cities (see Fig.~\ref{fig:exponents} and Table S1 in the SM).  
Fig.~\ref{fig:exponents} shows significant variability in the exponents $\beta$ and $z$, allowing us to distinguish broad city families according to their value of $\beta$ and $z$. We also show the values corresponding to the usual universality classes considered in the statistical physics of surface growth \cite{barabasi1995, odor2004} (see details in the SM). Along the $\beta$ axis, we can distinguish three broad families. The first group has low values, $\beta = 0.05 \pm 0.03$, indicating weak interface reactivity to population growth. These may signal a saturation regime, supported by small $1/z$ values suggesting halted correlation length growth. Alternative explanations include LCC densification (implying minimal area and roughness increase) or the absorption of similarly rough clusters. These cities generally experience low demographic pressure (except Atlanta). Within this group, $z$ is large and highly variable ($1.9$–$5$). The second group, with $\beta \approx 0.37 \pm 0.04$, aligns with MH, EW or KPZ universality classes, but $z$ ranges broadly and uniformly from $1.4$ to $3.7$. The third group has $\beta > 0.5$ and more constrained $z$ values ($1.25 < z < 1.43$), except for São Paulo ($z \approx 2.5$). These cities show faster roughness growth, likely due to high demographic pressure—though Paris and Santiago are exceptions. 
%
%
High $\beta$ values may result from anisotropy and quenched noise, leading to local pinning and directional growth. 
Overall, no city fits neatly into neither classical (EW, KPZ, MH), nor quenched (qMH, qKPZ) universality classes. This could have been expected given the complex interplay of local growth, coalescence and anisotropy of urban dynamics. Moreover, quenched noise affects importantly the behavior of growing interfaces \cite{barabasi1995, odor2004, costanza2003} as i) their characteristic exponents change with respect to their annealed counterpart, with typically rougher surfaces,  ii) breaks symmetries of the system (e.g Galilean invariance $\alpha+z=2$) and iii) introduces pinning-depinning transitions, at which interfaces progress between pinned phases via avalanches, also modifying the exponents with respect to the moving interface phase (far from the depinning transition).

{\it Conclusion---} Through this quantitative analysis of urban sprawl, we identified key patterns. Built-up area generally scales linearly with population, with density breakpoints marking transitions between densification and expansion regimes. Anisotropy can lead to branch-like growth, deviating from the average $r \sim \sqrt{P}$ behavior. The expansion of the largest connected component depends on demographic pressure: low pressure yields smooth, diffusion-like growth, while high pressure leads to coalescence-driven dynamics. Cities exhibit intrinsic anomalous roughening, quantified by an universal local roughness exponent $\alpha_{\text{loc}} \approx 0.5$, suggesting similar mechanisms across cities. The growth exponent $\beta$ (and $z$) varies widely, indicating that none of the standard growth equations apply, but instead revealing three distinct city families classified by the value of $\beta$. These results show that both local growth and coalescence are required to accurately describe urban sprawl, with additional crucial impact from anisotropy and quenched noise. 

Beyond scaling exponents, the full distribution of fluctuations is also of interest. The normalized radius does not follow a Tracy–Widom distribution, as observed in KPZ-type growth, but instead appears to follow a stretched exponential with an exponent close to $1/2$ (see Section 13 for details). Further investigation is needed to clarify the origin of this behavior.

This empirical characterization provides a necessary foundation for the mathematical modeling of urban sprawl: any theoretical model must reproduce these findings, which serve as essential benchmarks and guides for understanding and simulating urban expansion.

\section*{Acknowledgements}

U.M. and M.B. thank Maximilien Bernard and Alberto Rosso for enlightening discussions and suggestions about the $\alpha_{\text{loc}}$ value. M.B. expresses gratitude to Henri Berestycki for many insightful discussions on this topic over the years. R.G. thanks Francesca Bovolo for useful discussions.

\section*{Funding}

O. A. acknowledges support
from the Spanish Grants No. PID2021-128005NB-C22 and No. PID2024-158120NB-C22, funded by MCIN/AEI/10.13039/501100011033 and ``ERDF A way of making Europe'', from Generalitat de Catalunya (2021SGR00856).
R.G. acknowledges the financial support
received from the PNRR ICSC National Research
Centre for High Performance Computing, Big Data and Quantum Computing
(CN00000013), under the NRRP MUR program funded by the NextGenerationEU. \\

\section*{Authors contributions}

U.M. and R.G. collected the data. U.M., O.A., and R.G. processed the data. U.M. and O.A. developed the visualization code. U.M., R.G., and M.B. designed the study. U.M. and M.B. analyzed and interpreted the data and wrote the original draft. All authors interpreted and discussed the results, and reviewed and edited the manuscript.

\section*{Data availability statement}

All the datasets used in this work are publicly available \cite{worldpop, wsfevo, atlasurbanexpansion}.

\section*{Competing interests}
None. 

\section*{Additional information}

Supplemental Material is available for this paper \cite{Supplemental}.

\end{document}


\title{Supplemental Material for:\\
Universal roughness and\\ the dynamics of urban expansion}

\author{Ulysse Marquis$^{1,2}$}

\author{Oriol Artime$^{3,4}$}

\author{Riccardo Gallotti$^1$}

\author{Marc Barthelemy$^{5,6}$}
\email{marc.barthelemy@ipht.fr}

\affiliation{$^1$ Fondazione Bruno Kessler, Via Sommarive 18, 38123 Povo (TN), Italy}
\affiliation{$^2$ Department of Mathematics, University of Trento, Via Sommarive 14, 38123 Povo (TN), Italy}
\affiliation{$^3$Departament de Física de la Matèria Condensada, Universitat de Barcelona, 08028 Barcelona, Spain}
\affiliation{$^4$Universitat de Barcelona Institute of Complex Systems (UBICS), Universitat de Barcelona, 08028 Barcelona, Spain}
\affiliation{$^5$ Universit\'e Paris-Saclay, CNRS, CEA, Institut de Physique Th\'eorique, 91191, Gif-sur-Yvette, France}
\affiliation{$^6$ Centre d’Analyse et de Math\'ematique Sociales (CNRS/EHESS) 54 Avenue de Raspail, 75006 Paris, France}

\maketitle


\section{Data} \label{app:a}

The urban settlement data comes from the World Settlement Footprint Evolution dataset [28]. The data is a binary mask composed of pixels of size $30\text{m}\times30\text{m}$, for each year between 1985 and 2015. The built areas correspond to the value of pixel $1$. For countries part of the OECD, we extracted urban areas corresponding to the Functional Urban Areas (FUAs) [57]. For the cities in China, we extracted urban areas using FUAs provided by [58]. For the other cities, we manually selected boundaries surrounding the cities. For each city, we decomposed the built areas in connected clusters (based on their Moore neighborhood). We focused on the largest connected component (LCC) and define the city's center as the center of mass of the initial LCC. We define the aggregate quantities $A_{\text{tot}}$ and $A$ as the total built area and the built area of the largest cluster. The LCC are converted in centered radial profiles. $N_{\text{pts }}= 10^4$ points are sampled uniformly along each interface. For each $\theta_i = \frac{2\pi i}{N_{\text{pts}}}$, we set $r(\theta_i, t)$ as the distance to the furthest point belonging to the LCC at time $t$ making an angle $\theta_i$ with a horizontal plane passing through the center.

Population data is freely available on the websites Macrotrends and World Population Review [29]. The population is given for urban areas, we assume a homogeneous population distribution and write
$$
P = A \frac{P_{\text{tot}}}{A_{\text{tot}}}
$$
at each time $t$ ($P$ is the population of the LCC and $P_{\text{tot}}$ the population of the whole urban area).

The population and urban area data used in the analysis (section \ref{sec:atlas}) are freely available from [30].

The cities from the dataset [28] that were analyzed were chosen according to the following criteria :
\begin{itemize}
    \item They have large enough spatial extent for the angular profiles to be fine-grained
    \item Their population and area display substantial (positive) variation
    \item They don't display strong topographic constraints (e.g New York City)
    \item The yearly population data is available for the  corresponding urban area.
\end{itemize}
Concerning the cities from the dataset [30], no selection was made.

\section{Anisotropy exponent} 
\label{app:b}

Given $r(\theta, P)$ the radial profile at population $P$, we take $N$ sectors such that $N \delta \theta = 2 \pi$ and compute $\bar{r}_i(P) = \langle r(\theta, P)\rangle_i$, where $\langle.\rangle_i$ denotes the average over angle in $[\theta_i, \theta_i+\delta \theta]$. We then perform power-law fits of the form
\begin{equation}
    \bar{r}_i(P) \sim P^{\mu(\theta)},
\end{equation}
which results in a set of $N$ exponents that describes the growth as a function of population in different directions. In the isotropic case, the area scales as the population, leading  to $r \sim P^{1/2}$. Therefore, we compute the relative dispersion of the set $\{ \mu(\theta) \}$ with respect to $1/2$
\begin{equation}
    \text{Disp} = \frac{\sqrt{ \frac{1}{N}\sum_{i=1}^N  (\mu(\theta_i)-1/2)^2}}{1/2}.
\end{equation}
This measure quantifies the anisotropy of the growth of a city.

\section{Growth mechanisms} \label{app:c}

We define the aggregates on the LCC at time $t$ as the newly added connected components that form the LCC at time $t+1$, expressed as:  
\[
\text{Agg}(t+1) = \text{LCC}(t+1) \setminus \text{LCC}(t)
\]
These aggregates can be decomposed into connected clusters $c_1, c_2, \dots $. Each cluster $c$ of this set is composed of sites already built at time $t$ ($c_o(i))$ and of new sites built between time $t$ and $t+1$ ($c_n(i)$). Hence we can write $|c_i|=|c_o(i)|+|c_p(i)|$. Summing over the clusters
\begin{equation}
    C = \sum_i c_i = \sum c_o(i) + \sum c_n(i) = C_o + C_n
\end{equation}
The quantity $C_o$ quantifies the amount of coalescence while $C_n$ quantifies the amount of local growth. The temporal average $\langle (C_o+C_n)/A\rangle $ thus quantities the total growth (relatively to the existing area $A$), and the ratio $\langle C_0/C_n\rangle $ quantifies the importance of coalescence compared to local growth.

\section{Scaling function} \label{app:d}

The first dynamic scaling ansatz proposed in surface roughening theory is the Family-Vicsek one [59]
\begin{equation}
    w(L, t) = t^{\alpha/z} F(L t^{-1/z})
\end{equation}
with the scaling function obeying $F(u \ll 1) \sim u^\alpha$ and $F(u \gg 1) \sim 1$. Here $\alpha$ is the so-called \textit{roughness} exponent, $z$ the \textit{dynamic} exponent and $\beta=\alpha/z$ the \textit{growth} exponent. $L$ is the linear  size of the system. In this framework, there are only 2 independent exponents. 

Defining the \textit{local roughness} exponent $\alpha_{\text{loc}}$ through the relation $w(l,t) \sim l^{\alpha_{\text{loc}}}$ for $l \ll L$, we identify \textit{anomalous roughening} [52] when $\alpha_{\text{loc}} \neq \alpha$. This condition indicates that microscopic roughness scales differently from system-size roughness, deviating from standard roughening behavior. In this case, the Family-Vicsek is not valid anymore [52,60] and a more general dynamic scaling ansatz has been proposed [60]
\begin{equation} \label{eq:genericscaling}
    w(l,t) = t^\beta g(l t^{-1/z})\sim \left\{ 
    \begin{array}{ll}
        t^\beta  & \mbox{ if } t^{1/z} \ll l \\
        l^{\alpha_{\text{loc}}} t^{\beta^*} & \mbox{ if } l \ll t^{1/z} \ll L \\
        l^{\alpha_{\text{loc}}} L^{z \beta^*} & \mbox{ if } l \ll L \ll t^{1/z} 
    \end{array} \right.
\end{equation} 
where $\beta^* = \beta - \alpha_{\text{loc}}/z $ is nonzero whenever there is anomalous scaling and the \textit{correlation length} $\xi \sim t^{1/z}$ and $g(u \ll 1) \sim u^{\alpha_{\text{loc}}}$. When this scaling is valid, there are 3 independent exponents, in contrast with the usual Family-Vicsek scaling.

\section{Universality classes in kinetic roughening}

By considering locally normal growth of interfaces, Kardar, Parisi and Zhang introduced in 1986 [61] a general equation for surface growth which can be written in $1+1$ dimension
\begin{equation} \label{eq:kpz}
    \partial_t h = \nu \partial_{xx} h + \lambda (\partial_x h)^2 + \eta
\end{equation}
with $\langle \eta \rangle = 0$ and $\langle \eta(x,t) \eta(x',t') \rangle = 2 D \delta(x-x') \delta(t-t')$. The simplest case corresponds to $\lambda = 0$ for which there is no non-linear term and the equation corresponds to the Edwards-Wilkinson (EW) equation [21], whose characteristic exponents $\alpha = 1/2$, $\beta=1/4$ and $z = 2$ can be found by dimensional analysis. When $\lambda \neq 0$, 
the naive scaling analysis leads to wrong exponents -- $\nu$, $\lambda$ and $D$ are indeed coupled. A dynamic renormalization group analysis allows to find $\alpha=1/2$, $\beta=1/3$ and $z=3/2$. In 2010, Calabrese et al. [62] (among other groups), through a connection with directed polymers in random media, showed that solutions of the KPZ equation do have universal fluctuations, characterized by Tracy-Widom family of distributions. 

In Eq.~\ref{eq:kpz}, all the terms $\partial_x^{(2k)}h$ and $(\partial_x h)^{2k}$ with $k>1$ are neglected. Instead, the Mullins-Herring (MH) equation [63] 
\begin{equation} \label{eq:mh}
    \partial_t h = - \kappa \partial_{x}^{(4)} h + F_0 + \eta
\end{equation}
where $\kappa$ is the surface diffusion coefficient and $\eta$ is as in Eq.~\ref{eq:kpz}, used to describe molecular beam epitaxy, accounts for a fourth-order derivative. The MH equation is linear and can be solved using Fourier transform. Its characteristic exponents are $\alpha=3/2$, $\alpha_{\text{loc}}=1$, $z=4$, $\beta = 3/8$ and $\beta^*=1/8$ in $1+1$ dimensions. More classes exist, characterized by to a) conservative or non-conservative deterministic part b) linear or non-linear system c) conservative or non-conservative noise (see for example chapter 27 of [21] and [22]).

In all the previous cases, the noise $\eta$ was annealed : it resets at each instant. The quenched KPZ (QKPZ) equation describes the progression of interfaces through a random media 
\begin{equation} \label{eq:qkpz}
    \partial_t h = \nu \partial_{xx} h + \lambda (\partial_x h)^2 + F + \eta
\end{equation}
where $F$ is an external force, and the noise doesn't depend on time: $\langle \eta(x,h) \rangle = 0$ and $\langle \eta(x,h), \eta(x',h') \rangle =   \delta(x-x') \Delta(h-h')$. In general, QKPZ interfaces go through pinning-depinning transitions : there is a critical force $F_c$ which determines whether the interface moves or not, and how it moves : at $F \sim F_c$, the motion of the interface is not uniform, as it jumps between locally pinned phases via avalanches. Amaral et al [64] found that discrete models for surface growth in random media fell in general in two distinct universality depending on the behavior of $\lambda$ in function of $f = \frac{F-F_c}{F_c}$. In the $\lambda \to 0$ case, Eq.~\ref{eq:qkpz} can also be seen as the quenched EW equation [22]. 
When $\lambda > 0$, the QKPZ equation seems to be connected with directed percolation (DP) at the depinning transition. Exponents $\alpha \approx 0.63$, $\beta \approx 0.63$, $z\approx1$ are found (at the depinning transition). 

Similarly, the quenched MH equation was studied by Lee and Kim [65] and has $\alpha \approx1.50$, $\beta \approx 0.84$ and $z \approx 1.78$ at the depinning transition.
 It is worthy of noting that at the moving phase $F \gg F_c$, different exponents characterize the exponents (some experiments even suggest crossovers for values of exponents [21]). In general, interfaces growing through random media do show high roughness and growth exponents, $\alpha$, $\beta$ and can violate the Galilean invariance equality $\alpha+z=2$.

\section{Exponent measures} \label{app:e}

In the context of city growth, the interface is two-dimensional and anisotropic, requiring adaptations to standard exponent measurements, which are typically defined for one-dimensional or isotropic cases. 
At variation with the $1+1$ case with band geometry, the definition of the system size is not straightforward. If we consider the perimeter as a measure of size, its value is non-trivial due to the fractal nature of the urban giant component’s boundary [6]. An alternative approach is to use the average perimeter, defined as $2\pi \bar{r}$, where 
$\bar{r}$ is the average radius. However, this measure loses relevance due to the anisotropic nature of urban expansion—it is not even guaranteed to increase as $A$ grows. Furthermore, the perimeter varies over time and lacks a fixed characteristic value, making it an inconsistent metric for system size. This issue complicates the measure of $\alpha$ ; other techniques based on the computation of the power spectrum [52,60,47] and two-point functions [48] allow for the determination of $\alpha$ in the band geometry [52,60] and in the radial geometry cases [47,48] but are difficult to extend to the anisotropic case.


\subsection{Measure of $\alpha_{\text{loc}}$}

We first recall the definition of $\alpha_{\text{loc}}$
\begin{equation}
    w(l \ll t^{1/z}) \sim l^{\alpha_{\text{loc}}} 
\end{equation}
at fixed $t$. Hence, by splitting the interface in $N$ sectors of aperture $\Delta \theta$ and taking $\bar{\ell} = 1/N \sum r(\theta) \Delta \theta$ where $r(\theta)$ is the average radius of the sector, we can measure the local width as in Eq.~2 in the main text. Varying $\Delta \theta$, we obtain couples $(\bar{\ell}, w(\bar{\ell}))$ and we compute $\alpha_{\text{loc}}$ by using a linear regression in log-log scale.

\subsection{Measure of $z$}

As discussed earlier, direct measurements of $z$ are not feasible in this setup. Instead, we estimate $1/z$ by first measuring the correlation length $\xi$ of the interface and then using the scaling relation 
$\xi \sim P^{-1/z}$. To compute $\xi$, we extend the \textit{patches} method introduced in [66]. In the isotropic (circular) case, the procedure is as follows: we determine the average radius $\bar{r}$ of the interface and partition it into $n_P$ patches, each corresponding to successive segments of the interface that are either above or below $\bar{r}$. The correlation length is then defined as the expected length of the patch to which a randomly chosen point belongs, given by:
\[
\xi \approx \frac{\sum l_i^2}{\sum l_i}
\]
where \( l_1, \dots, l_{n_P} \) are the patch lengths.

To account for anisotropy, we extend this method by introducing localized radii. Specifically, we divide the interface into $N$ angular sectors of width $\Delta \theta$ and compute local average radii $\bar{r}_i = \langle r(\theta) \rangle_i$. This refinement ensures that the correlation length measurement remains meaningful despite directional variations in growth.

Using this approach, there are at least $2N$ patches, which naturally sets an upper bound on $\xi$. However, by maintaining a moderate number of sectors and assuming that typical local radii grow faster than the correlation length, the method should still reliably capture the growth of $\xi$.

\subsection{Measure of $z$ and $\beta$}

Another way to measure $1/z$ and $\beta$ is by optimizing the collapse $w P^{-\beta} \sim (\bar{\ell} P^{-1/z})^{\alpha_{loc}}$ (assuming the regime is $l \ll t^{1/z}$). There is a curve for each of the $T$ times of observation. If for each curve there are $M$ points $(x_{i,t}, y_{i,t}) = (\log(\bar{\ell_t}P_t^{-1/z}), \log(w(\bar{\ell})P_t^{-\beta}))$, we minimize over $\beta$ and $z$ the `thickness' of these curves, given by
\begin{equation}
    \sum_{k=1}^M \sum_{1 \leq i < j \leq T} || \  (x_{k,i}, y_{k,i}) - (x_{k,j}, y_{k,j}) \ ||^2.
\end{equation}

\subsection{Measure of $\beta^*$}

Similarly to the technique exposed in the previous section, a way to measure $\beta^*$ is to optimize the collapse $w P^{-\beta^*} \sim \bar{\ell}^{\alpha_{\text{loc}}} $ in the same regime. This amounts here to minimize the gap (measured on the $y$-axis) between points with the same abscissa.

\section{Analysis for the period 1800-2000} \label{sec:atlas}


The dataset from [30] includes records of urban area and population spanning a 200-year period (1800–2000). With data points spaced approximately 29 years apart on average, the analysis of $A_{\text{tot}}$ versus $P_{\text{tot}}$ remains relatively coarse-grained. Additionally, the methodology used to generate the dataset may introduce some inconsistencies, such as instances where the built area decreases despite population growth. However, the dataset’s extensive temporal coverage and geographical diversity make it a valuable and robust resource for studying urban area expansion over time. 

\begin{figure}[htp]
    \centering
    \includegraphics[width=0.99\linewidth]{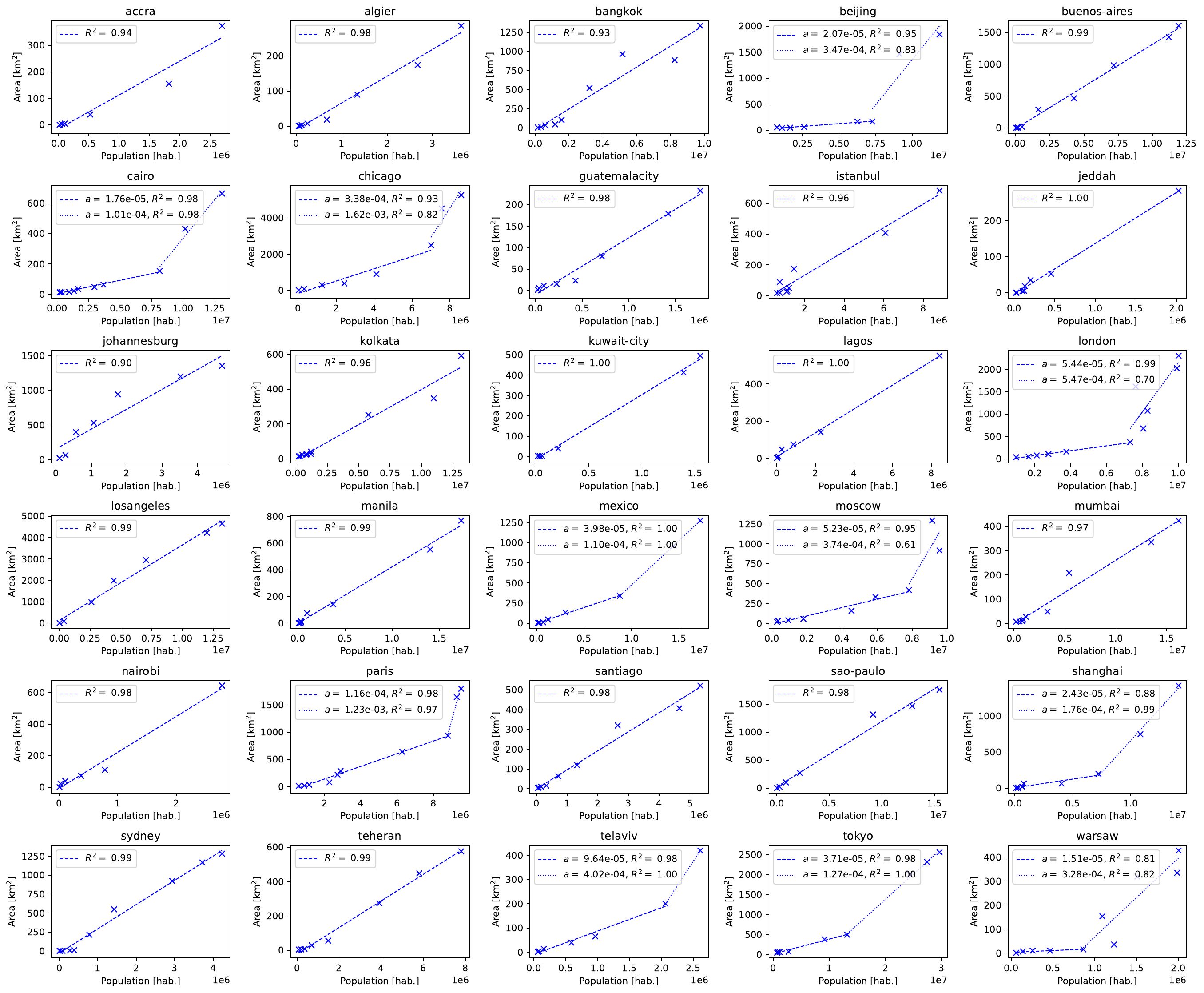}
    \caption{\textbf{Linear approximation of core area against population for the cities of the dataset [30] (period 1800-2000).} In 20 cities over the 30 studied, cities grow at constant density, corresponding to a single slope, over the period studied (1800-2000). In the other cases, there is a density breaking point, for which in all cases the population density decreases, sign of spatial expansion of the cities. }
    \label{fig:prelim-longitudinal}
\end{figure}

We begin by plotting the longitudinal evolution, showing how the area $A_{\text{tot}}$ changes with population $P_{\text{tot}}$ for each city over time. We observe that piecewise linear functions approximate well the built urban area when population grows, see Fig. \ref{fig:prelim-longitudinal}. We observe mostly 2 types of growth on the observed period: 
\begin{itemize}
    \item (i) Growth at constant density. For instance, for Los Angeles (USA) or Buenos Aires (Argentina).
    \item (ii) A density breaking point, separating a dense growth in early periods from a large expansion in later times (for example, Cairo (Egypt) or Shanghai (China)).
\end{itemize}

\begin{figure}[htp]
    \centering
    \includegraphics[width=0.5\linewidth]{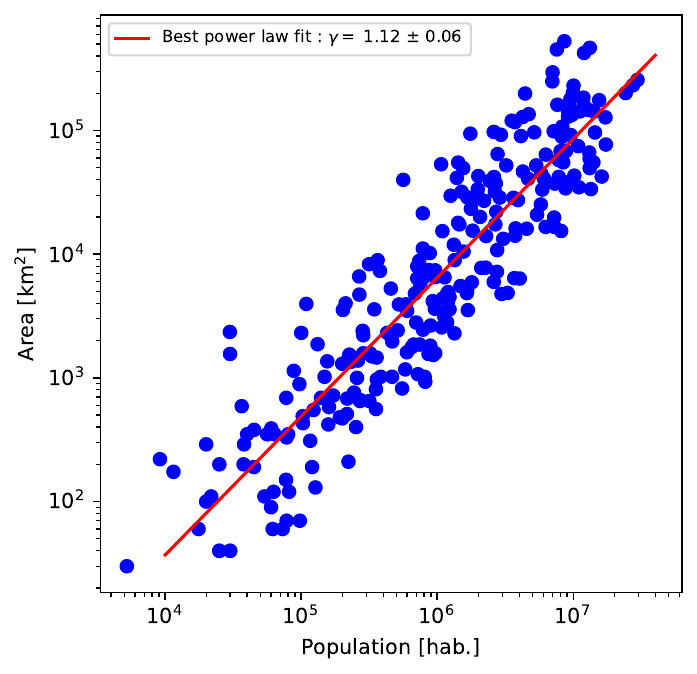}
    \caption{\textbf{Urban area versus population for the cities of the dataset [30] -- transversal scaling.} Each symbol corresponds to each city for one date. A power law fit reveals an exponent of $1.12\pm0.06$.}
    \label{fig:prelim-transversal}
\end{figure}

By aggregating data across all years and cities, we obtain the `cross-sectional' or `transversal' scaling relationship $A_{\text{tot}} \sim P_{\text{tot}}^{\gamma}$ and observe an exponent $\gamma = 1.12 \pm 0.06$ ($R^2 = 0.83$), as shown in Fig. \ref{fig:prelim-transversal}. This slight supra-linear trend suggests that population density, given by $ \rho = P_{\text{tot}}/A_{\text{tot}} \sim A_{\text{tot}}^{1-\gamma}$, decreases gradually as population increases.

\begin{figure}[htp]
    \centering
    \includegraphics[width=0.8\linewidth]{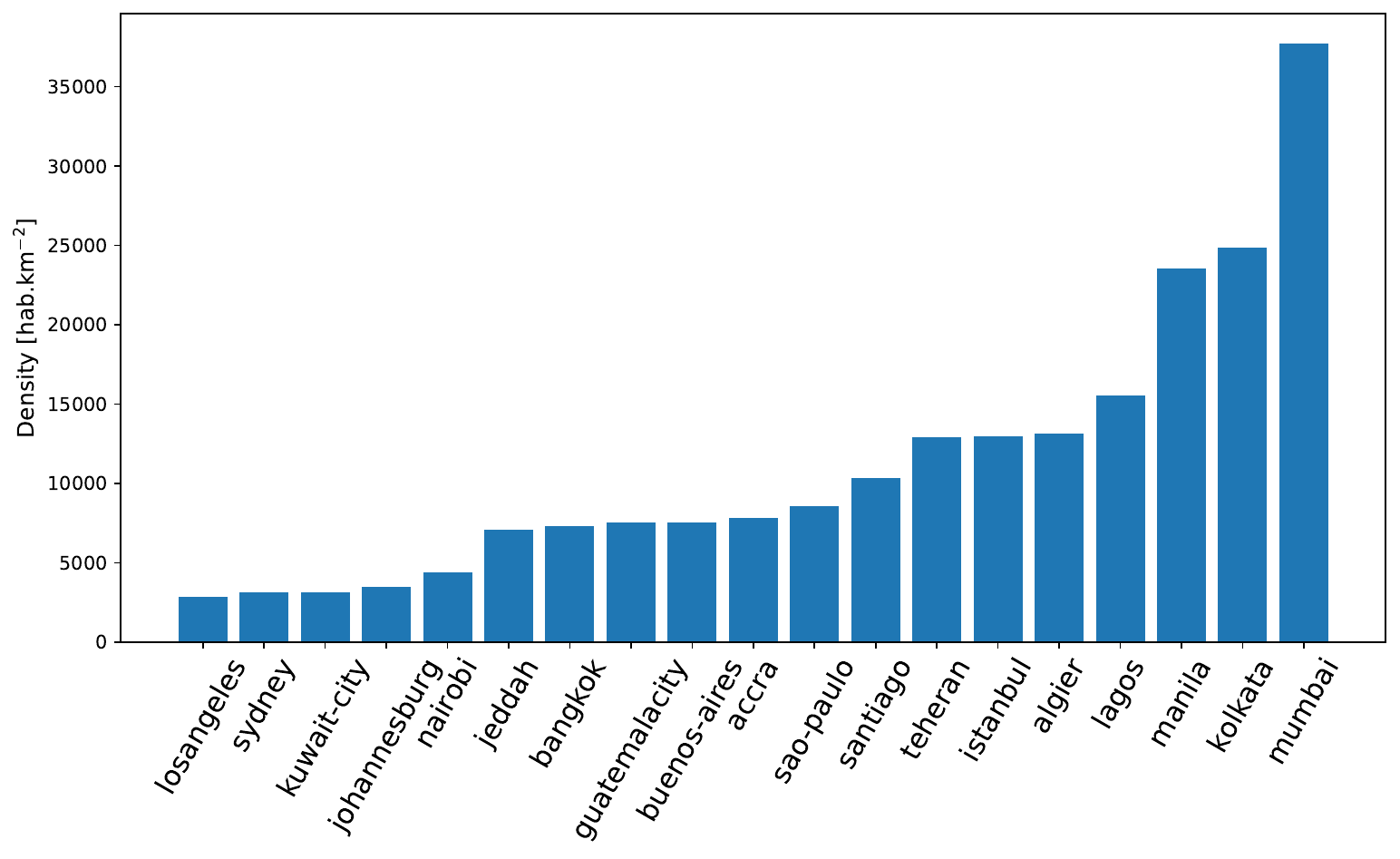}
    \caption{\textbf{Densities of population for the cities of the dataset [30] with one single growth regime.}}
    \label{fig:prelim-densities}
\end{figure}

For cities growing at constant rate, we also show the distribution of densities in Fig. \ref{fig:prelim-densities}. We observe large differences with 5 cities below $5,000$ inhabitants per km$^2$ and 3 cities above $20,000$.

\section{Area growth}

\begin{figure}[htp]
    \centering
    \includegraphics[width=0.99\linewidth]{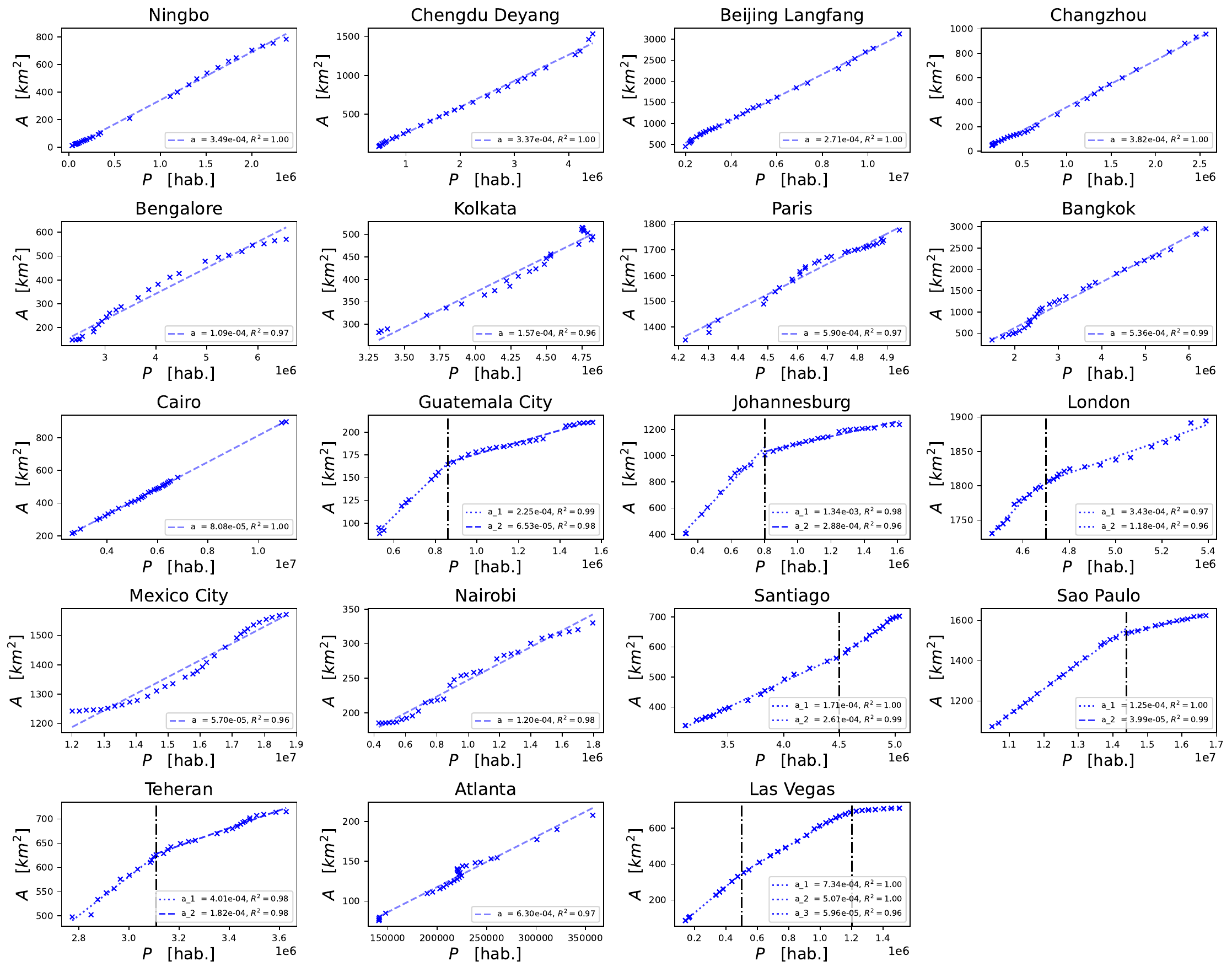}
    \caption{\textbf{Growth of the largest component of 19 urban areas.} The trends are approximated by piecewise linear functions, the change of slopes are marked by the black vertical lines. In the legend, we show the slopes, where $A \approx a P + b$, with $a$ representing essentially the inverse density. Densities of the LCC range in $[750; 25,000] \text{ hab. } \text{km}^{-2}$.
    Most cities (12 out of 19) are approximated by a single linear trend, corresponding to growth at constant density. 6 have two regimes, among which 5 have a smaller slope in the second regime, corresponding to a densification phase of the city (i.e. a growth at a larger density). Only Santiago has a slightly steeper slope in its second phase. Finally, Las Vegas possesses 3 phases, corresponding to 2 regimes of growth at different densities, and then a saturation regime where the city reached limits (in that case, boundaries imposed by federal land).}
    \label{fig:area_v_pop_lcc}
\end{figure}

\begin{figure}[htp]
    \centering
    \includegraphics[width=0.8\linewidth]{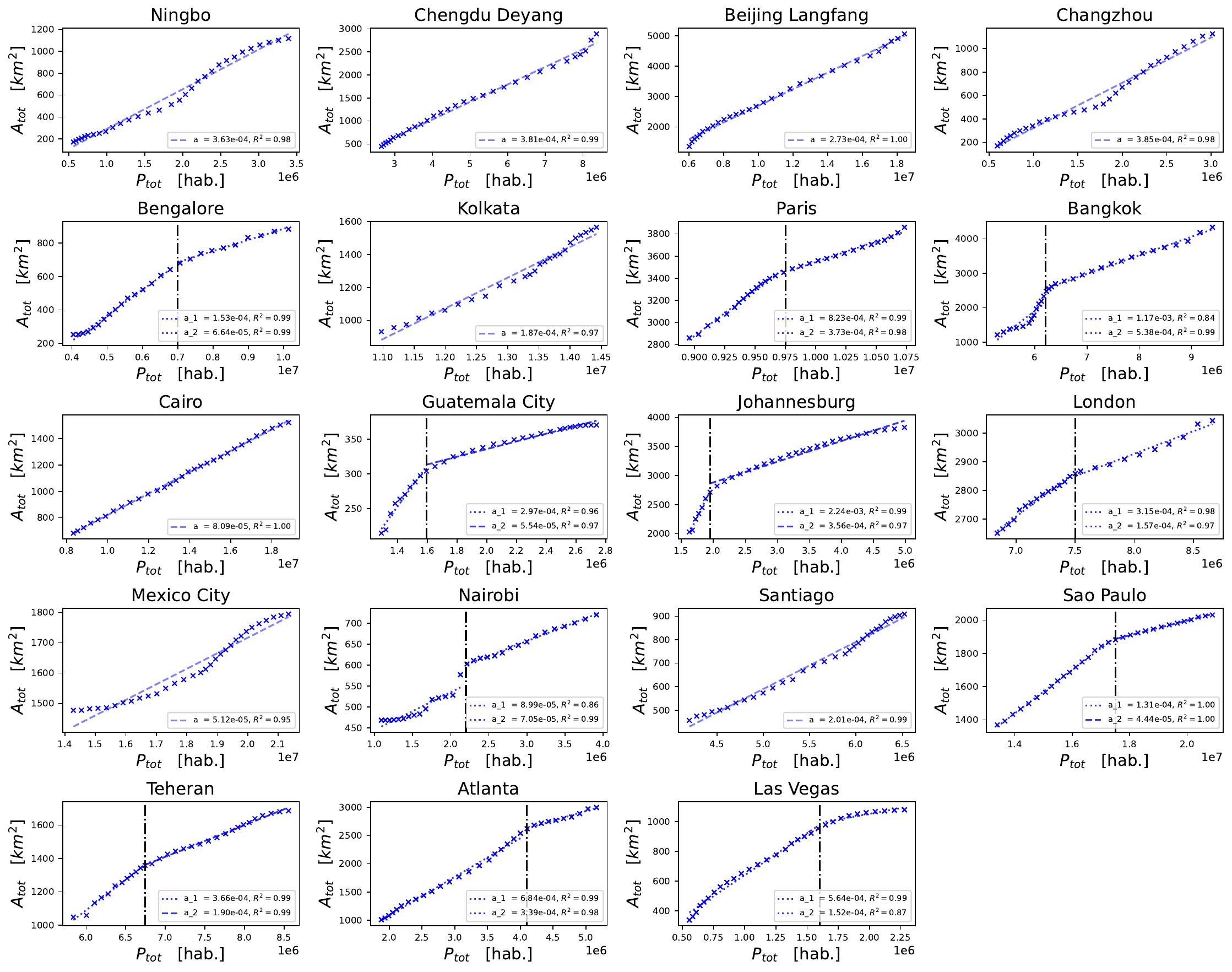}
    \caption{\textbf{Growth of the total built area.} Most cities are well approximated by piecewise linear functions, where the second regime has a smaller slope corresponding to a densification phase. Densities (measured as inverse of the slopes) range in $[450; 23,000]$  \text{hab.} \text{km}$^{-2}$}.
    \label{fig:area_v_pop_tot}
\end{figure}

The growth of the urban area versus the population, both for the largest connected component and the whole urban area, are shown in Fig. \ref{fig:area_v_pop_lcc} and \ref{fig:area_v_pop_tot}, respectively. In both cases, the built area is well approximated by piecewise linear functions of the population. However the breaking points do not happen necessarily at the same date across the two cases. This can be explained by the fact that the LCC and the urban area do not necessarily grow the same way, for instance the LCC might get pinned and stop growing while the surrounding clusters continue to develop (see, e.g.,Las Vegas). Deviations from the linear interpolation are found, for example for Mexico City or Nairobi. However, in both cases, the piecewise linear fits are robust, with the $R^2$ in $[0.96, 1.0]$ for the LCC and $[0.84, 1.0]$ for the whole urban area.


In Fig.~\ref{fig:area_v_pop_tot}, we present the results for the total built area, rather than just the LCC. The observed trends closely resemble those of the LCC, with piecewise linear functions providing a good fit to the data. 

\section{Anisotropy measure}

\subsection{Exponent histograms}

\begin{figure}[htp]
    \centering
    \includegraphics[width=0.8\linewidth]{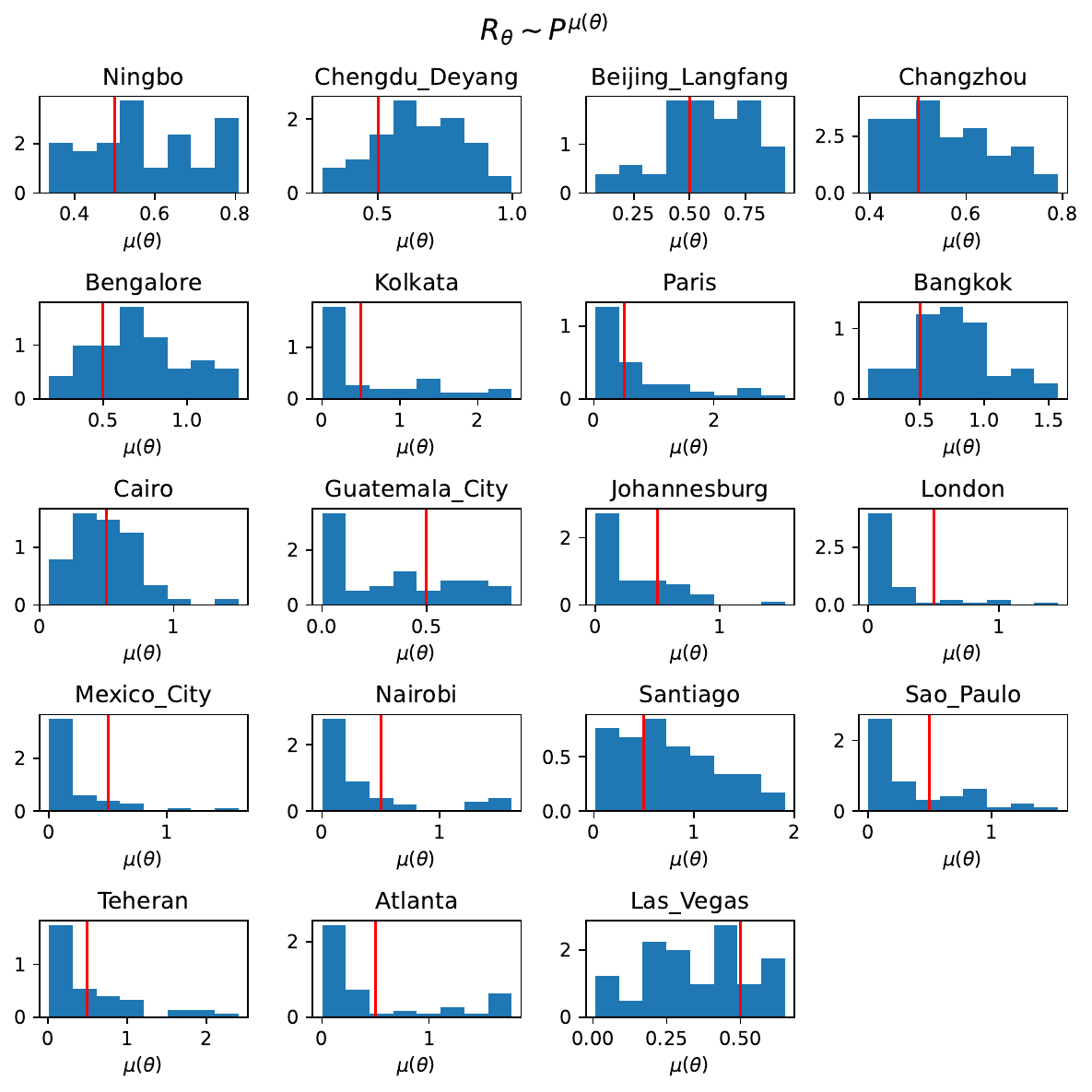}
    \caption{\textbf{Distribution of $\mu(\theta)$.} The red vertical line indicates the values $\mu=1/2$.}
    \label{fig:histo-anisotropy}
\end{figure}

The histogram of the exponents $\mu(\theta)$ is shown for each city in Fig. \ref{fig:histo-anisotropy}. 
Two broad families seem to emerge:
\begin{itemize}
    \item (i) The distribution well concentrated around the isotropic radial growth exponent $\mu = 1/2$. 
    \item (ii) The distributions with exponents mostly concentrated around $0$ with some exceptions, characterized by long tails going up to $\mu \approx 2$ in some cases. This could represent cities with most of its interface pinned ($\mu \approx 0$) and branches extending rapidly in some directions ($\mu >1$).
\end{itemize}

\subsection{Dispersion of exponents}

\begin{figure}[htp]
    \centering
    \includegraphics[width=0.8\linewidth]{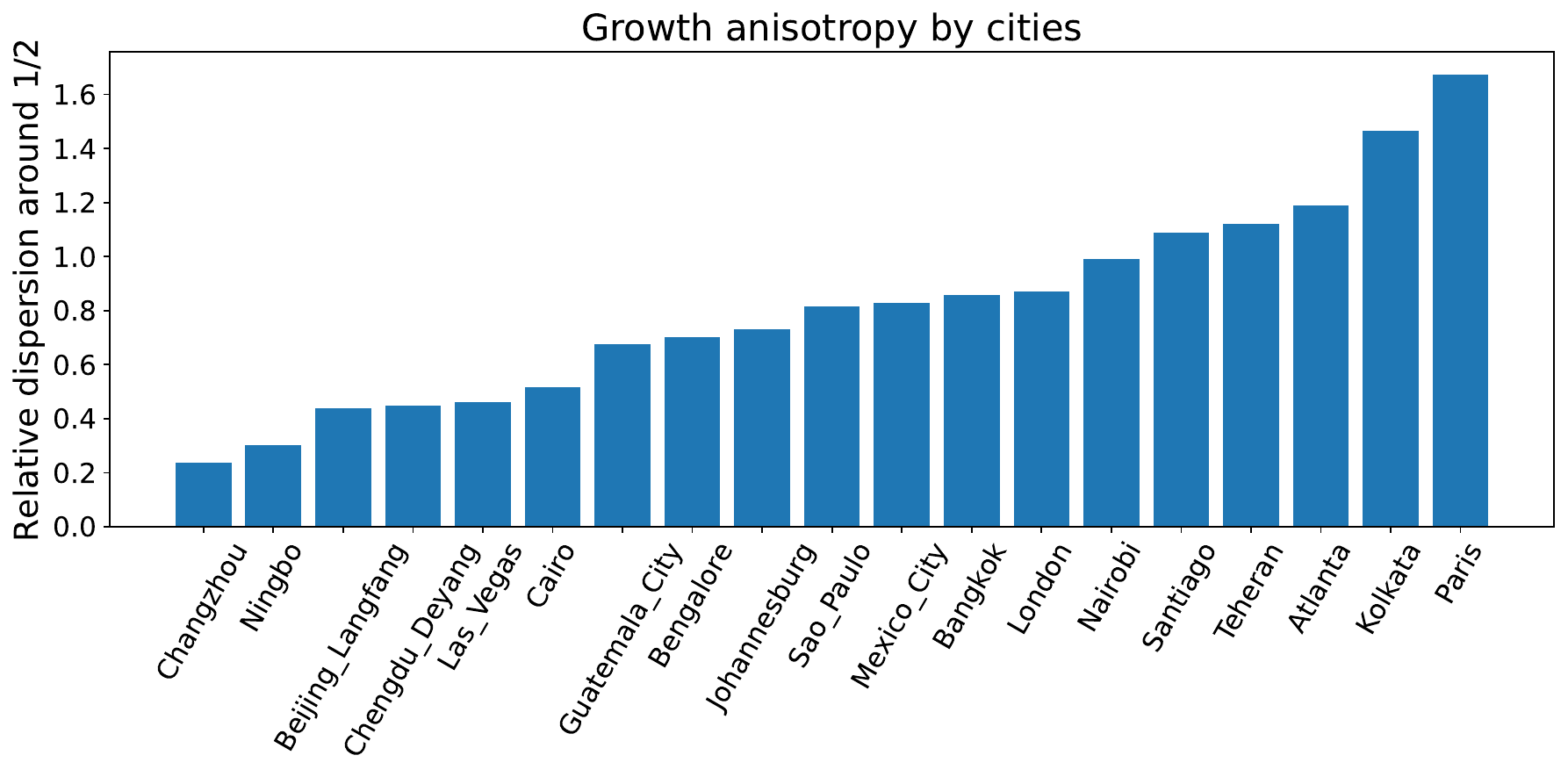}
    \caption{\textbf{Cities ranked by relative dispersion of $\mu(\theta)$ around $1/2$.}}
    \label{fig:barplot-dispersion}
\end{figure}

The relative dispersion of $\mu(\theta)$ around $1/2$ is shown in Fig. \ref{fig:barplot-dispersion} for the cities considered here. 

\begin{figure}[htp]
    \centering
    \includegraphics[width=0.8\linewidth]{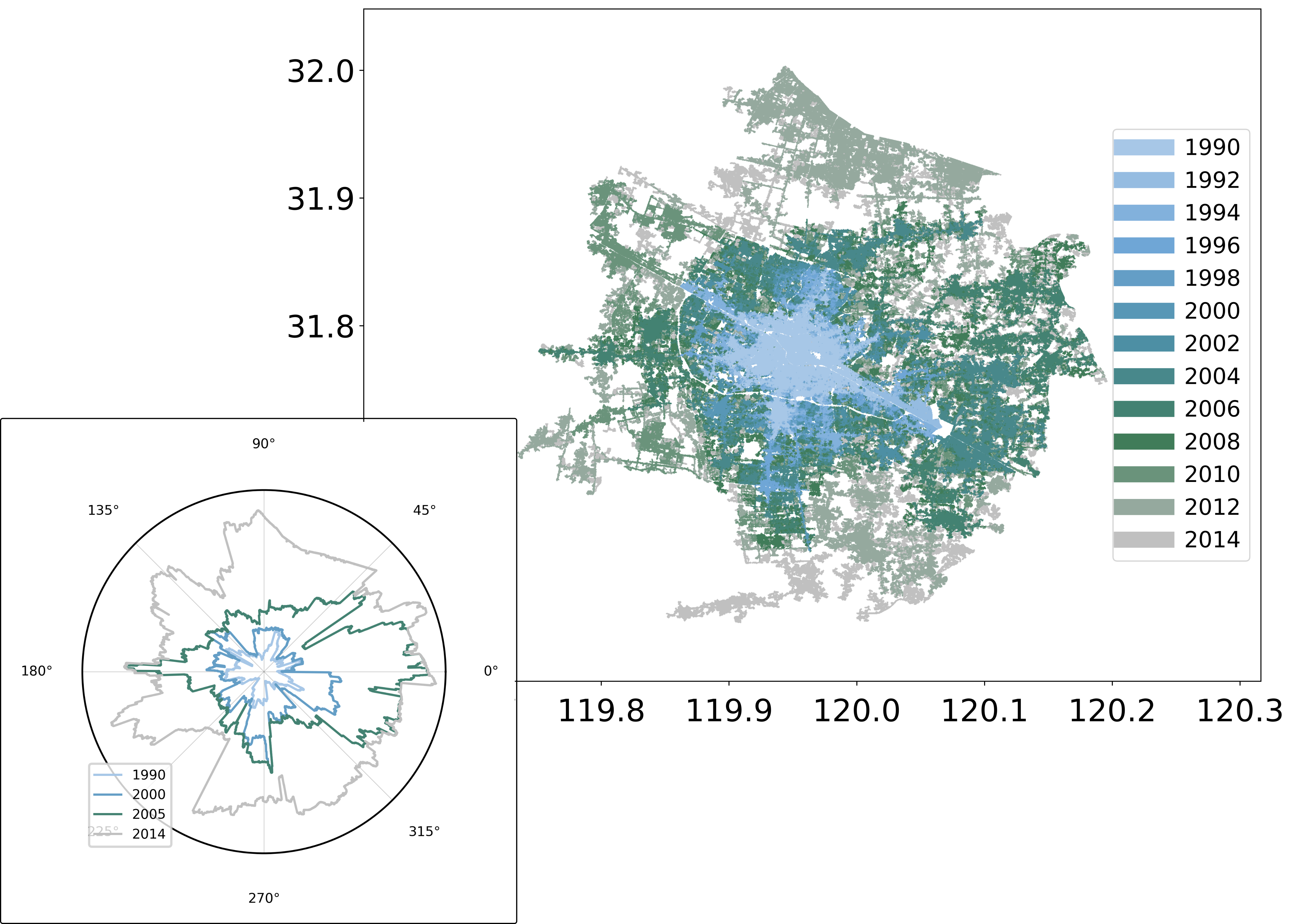}
    \caption{\textbf{Example of isotropic growth. Map of Changzhou (China).}}
    \label{fig:map-changzhou}
\end{figure}

\begin{figure}[htp]
    \centering
    \includegraphics[width=0.8\linewidth]{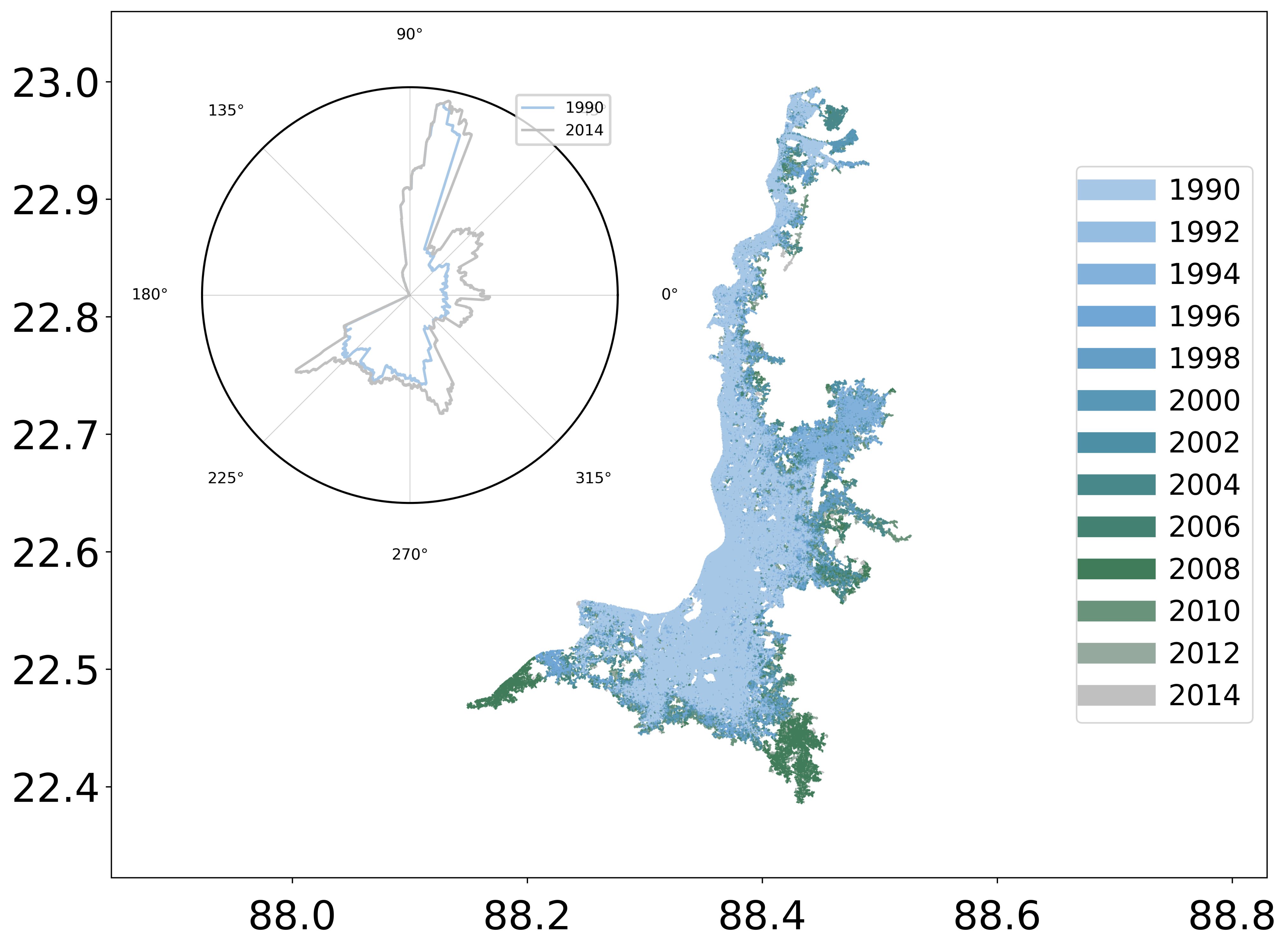}
    \caption{\textbf{Example of anisotropic growth. Map of Kolkata (India).}}
    \label{fig:map-kolkata}
\end{figure}

Fig. \ref{fig:map-changzhou} and \ref{fig:map-kolkata} Maps illustrate isotropic and anisotropic growth.

\section{Aggregates sizes}

\begin{figure}[htp]
    \centering
    \includegraphics[width=0.9\linewidth]{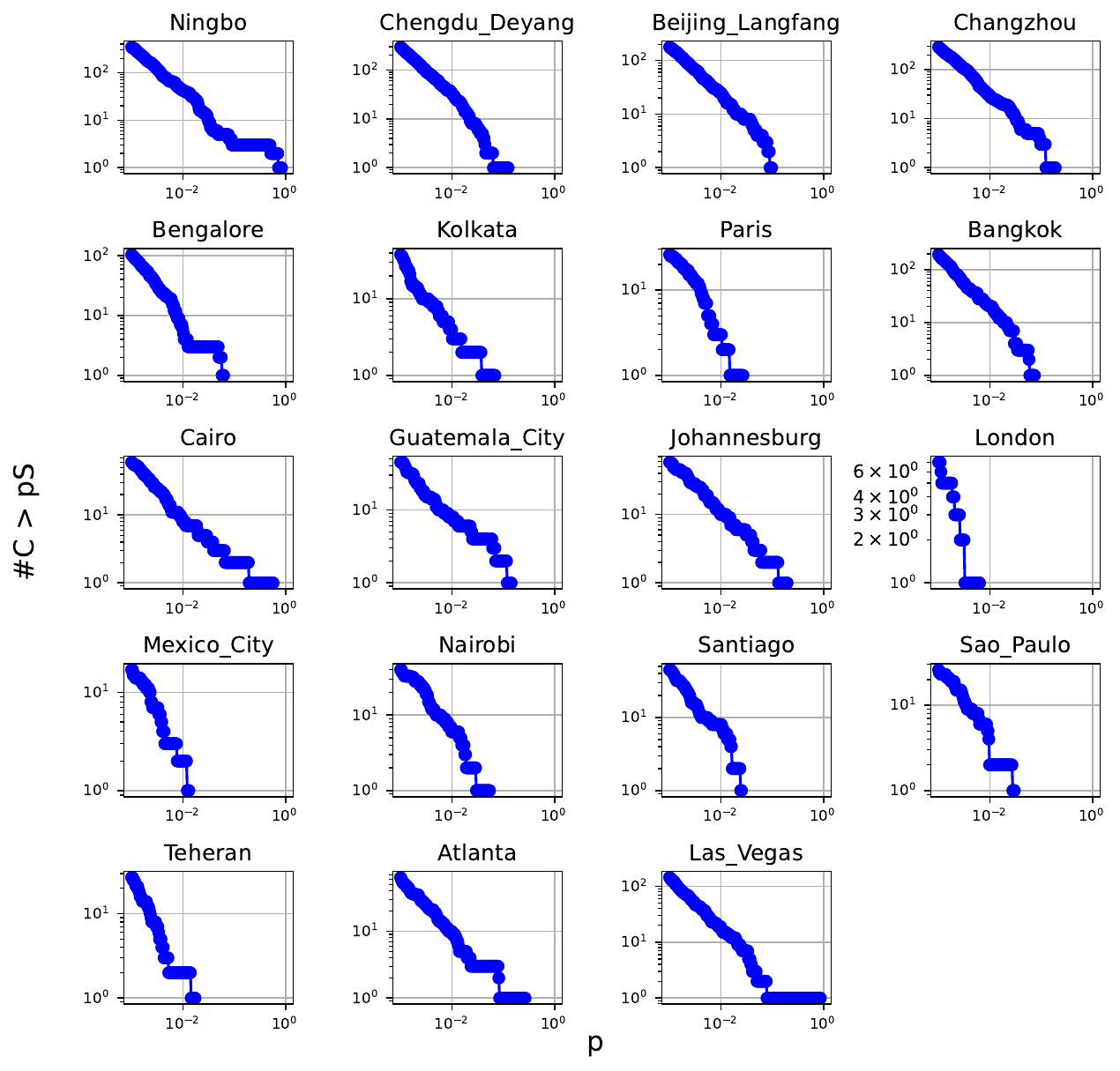}
    \caption{\textbf{Cumulated number of aggregates of size larger than $pS$.} For $p \in [10^{-3},1]$, we compute the number of aggregates of size larger than $pS$ and sum it over the years (plots shown in loglog).}
    \label{fig:enter-label}
\end{figure}

We show in Fig.~\ref{fig:enter-label} the probability to observe an aggregate (normalized by the size of the LCC) larger than a certain value.  We observe that the relative size of aggregates can be remarkably large. In cities such as Ningbo, Cairo, Atlanta, and Las Vegas, clusters comparable in size to the LCC are absorbed at certain points in time. This is reflected in the cumulative distribution, which exhibits nonzero values near 1. For these cities, coalescence is a crucial part of their growth process.

\section{Surface growth exponents}

Table \ref{tab:exponents} contains the measures of the exponents $ \alpha_{\text{loc}}$, $\beta$, $1/z$ and $\beta^*$. $\beta$, $1/z$ and $\beta^*$ are measured through the optimal collapse method explained in the Methods. $\beta^*$ can be compared to $\beta-\alpha_{\text{loc}}/z$. We also show the value of $\alpha = \beta z$. For the cities for which it was possible, we computed $1/z$ through the method of patches, described in the Methods, with $N_{sectors}=120$. To ensure consistency of the measures, we verified that the results were stable with little variations of $2\pi/N_{sectors}$. All the $R^2$ values of the linear fit in log-scale $\xi \sim P^{1/z}$ are above $0.90$, except for Bengalore for which $R^2 = 0.86$.

\section{A statistical explanation for $\alpha_\text{loc} \approx  1/2$}

\begin{figure}
    \centering
    \includegraphics[width=1\linewidth]{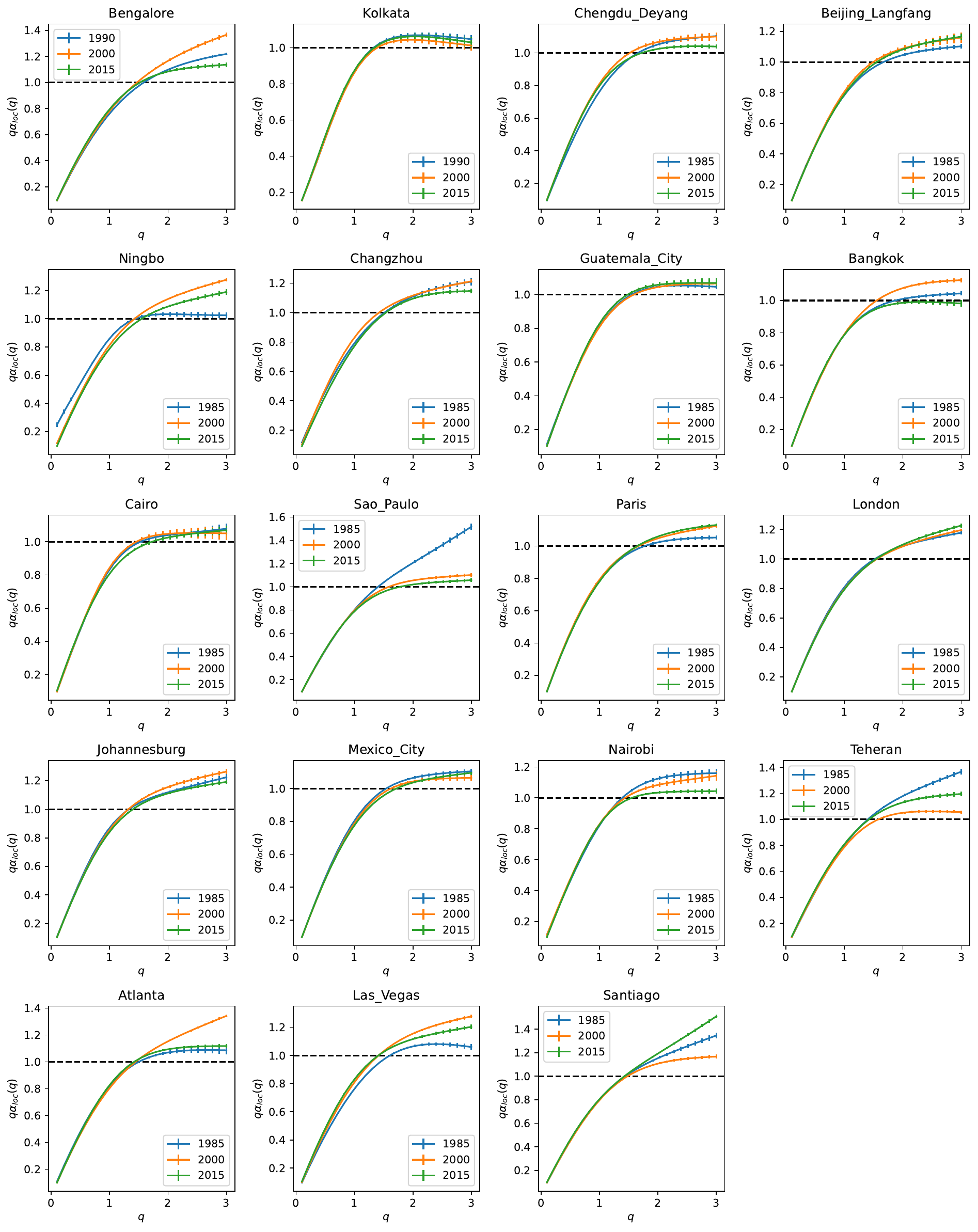}
    \caption{{\bf Plot of $q\alpha_{loc}(q)$ versus $q$ for different cities and different times.} If the origin of $\alpha_{\text{loc}}(q=2)$ is of purely probabilistic origin [55], we expect as in [54] a linear behavior for $q<q_c$ and a plateau at 1 for $q>q_c$. }
    \label{fig:qalphaloc}
\end{figure}

The result $\alpha_{\text{loc}}\approx 1/2$ could suggest a probabilistic origin as in [54,55]. The main ingredient is that the variance per sector is broadly distributed. This implies for the $q^{th}$ moment a behavior of the form 
$w^q\sim \ell^{\alpha_{\text{loc}}(q)}$ with [55]
\begin{align}
    \alpha_{\text{loc}}(q)q =1
\end{align}
In the model discussed by Bernard et al. [54], two regimes are expected for $q \alpha_{\text{loc}}(q)$ : a linear growth up until $q=q_c$, and a plateau $q \alpha_{\text{loc}}(q) = 1$ for $q>q_c$. These considerations motivate the inspection of the profiles $q \alpha_{\text{loc}}(q)$, that we show in Fig.~\ref{fig:qalphaloc}. In this Figure, we observe this behavior for some cities (Cairo, Kolkata, Chengdu, Guatemala City), but not for all cities. For example for cities such as Santiago, Atlanta, Las Vegas, etc. we don't observe a plateau for $q>q_c$. At this stage, the origin of $\alpha_{\text{loc}} \approx 1/2$ remains unclear. This issue warrants further investigation, with improved statistics to enable a reliable computation of higher moments of the interface width.

\section{Interface fluctuations}

In order to characterize the full distribution of fluctuations, it is essential to account for their intrinsically anomalous nature. In contrast to the standard Family–Vicsek scaling, local and global fluctuations do not exhibit the same scaling behavior. Specifically, for a city of population $P$, the local width scales as
\begin{align}
    w(\ell, P) \sim \ell^{\alpha_{\text{loc}}} P^{\beta^*}
\end{align}
when the observation length $\ell \ll \xi$, where $\xi$ denotes the correlation length. This relation enables a comparative analysis of fluctuations across cities of different sizes and at varying observation scales $\ell(\Delta\theta)$ by considering the rescaled variable
\begin{align}
    x = \frac{r - \langle r \rangle_{\Delta \theta}}{\ell^{\alpha_{\text{loc}}} P^{\beta^*}},
\end{align}
where $\langle \cdot \rangle_{\Delta \theta}$ denotes the average over an angular sector of size $\Delta\theta$, corresponding to a length scale $\ell(\Delta\theta)$. This approach is valid under the assumption that the correlation length $\xi$ is larger than the segment length at which the fluctuations are computed.

\begin{figure}
    \centering
    \includegraphics[width=1\linewidth]{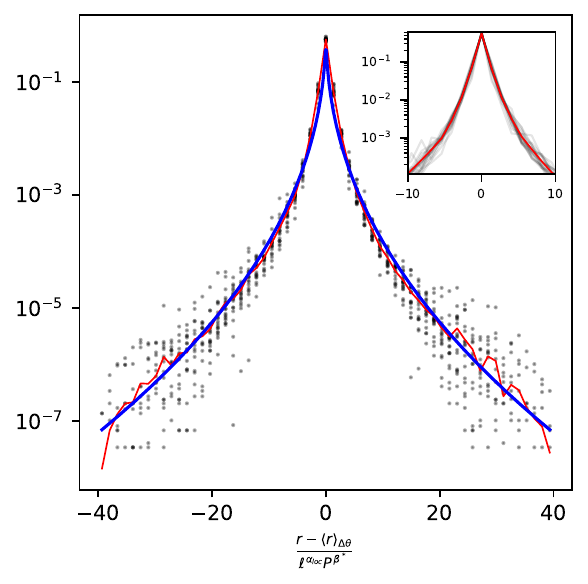}
    \caption{{\bf Standardized distribution of fluctuations across cities (grey points).} Red line : average profile. Blue line : best fit in $\log y = \log a - b x^c $ with $a = 0.97 \pm 0.37$, $b = 3.06 \pm 0.27$ and $c = 0.46 \pm 0.02$. Inset : zoom on $x \in [-10, 10]$}
    \label{fig:fluctuations}
\end{figure}

Standardizing the fluctuations in this way allows us to collapse the distributions across cities, which are expected to differ only by a multiplicative constant. In Fig.~\ref{fig:fluctuations}, we show the resulting standardized distributions for multiple cities, along with the corresponding average fluctuation profile. Notably, we observe a good collapse, but the rescaled distribution does not follow the Tracy–Widom form, characteristic of thermal KPZ surface growth, as commonly observed in other systems [16,17,67]. Instead, we observe a collapse onto a symmetric (differently from TW) stretched exponential distribution of the form
\begin{align}
    P(x) \sim \exp\left( -b |x|^c \right),
\end{align}
with fitted parameters $b = 3.06 \pm 0.27$ and $c \approx 1/2$ (the obtained fitting value is $c= 0.46 \pm 0.02$).

\section{Videos}

We provide in [68] videos of the growth of each urban area considered here for the period 1985 to 2015. 


\ulysse{
\begin{table}[htp]
    \centering
    \renewcommand{\arraystretch}{1.2}
    \setlength{\tabcolsep}{6pt}
    \begin{tabular}{|l|c|c|c|c|c|c|c|}
        \hline
        \textbf{City} & $\boldsymbol{\alpha_{\textbf{loc}}}$ & $\boldsymbol{\beta}$ & $\boldsymbol{1/z}$ & $\boldsymbol{\beta - \alpha_{\text{loc}}/z}$ & $\boldsymbol{\beta^*}$ & $\boldsymbol{1/z}$ (patches) & $\boldsymbol{\alpha = z \beta}$ \\
        \hline
        \textbf{Ningbo} & $0.56 \pm 0.03$ & 0.44 & 0.58 & $0.12 \pm 0.02$ & 0.10 & $0.52 \pm 0.05$ & 0.76 \\
        \textbf{Chengdu Deyang} & $0.53 \pm 0.02$ & 0.68 & 0.68 & $0.32 \pm 0.01$ & 0.31 & $0.63 \pm 0.07$ & 1.00 \\
        \textbf{Beijing Langfang} & $0.54 \pm 0.03$ & 0.41 & 0.58 & $0.10 \pm 0.02$ & 0.10 & $0.51 \pm 0.08$ & 0.71 \\
        \textbf{Changzhou} & $0.54 \pm 0.02$ & 0.37 & 0.56 & $0.07 \pm 0.01$ & 0.10 & $0.48 \pm 0.05$ & 0.66 \\
        \textbf{Bengalore} & $0.55 \pm 0.02$ & 0.83 & 0.72 & $0.43 \pm 0.01 $ & 0.47 & $0.60 \pm 0.15 $ & 1.15 \\
        \textbf{Kolkata} & $0.52 \pm 0.02$ & 0.34 & 0.74 & $-0.04 \pm 0.01$ & -0.02 & - & 0.46 \\
        \textbf{Paris} & $0.52 \pm 0.01$ & 0.56 & 0.76 & $0.16 \pm 0.01$ & 0.18 & - & 0.74 \\
        \textbf{Bangkok} & $0.53 \pm 0.02$ & 1.01 & 0.80 & $0.59 \pm 0.02$ & 0.59 & $0.74 \pm 0.10$ & 1.26 \\
        \textbf{Cairo} & $0.53 \pm 0.01$ & 0.37 & 0.54 & $0.08 \pm 0.01$ & 0.10 & - & 0.69 \\
        \textbf{Guatemala City} & $0.52 \pm 0.01$ & 0.37 & 0.33 & $0.20 \pm 0.00$ & 0.18 & - & 1.12 \\
        \textbf{Johannesburg} & $0.58 \pm 0.01$ & 0.07 & 0.27 & $-0.09 \pm 0.00$ & -0.06 & - & 0.26 \\
        \textbf{London} & $0.54 \pm 0.00$ & 0.01 & 0.21 & $-0.10 \pm 0.00$ & -0.06 & - & 0.05 \\
        \textbf{Mexico City} & $0.55 \pm 0.01$ & 0.04 & 0.25 & $-0.10 \pm 0.00$ & -0.06 & - & 0.16 \\
        \textbf{Nairobi} & $0.56 \pm 0.02$ & 0.28 & 0.27 & $0.13 \pm 0.01$ & 0.14 & - & 1.04 \\
        \textbf{Santiago} & $0.58 \pm 0.02$ & 0.62 & 0.74 & $0.19 \pm 0.01$ & 0.18 & - & 0.84 \\
        \textbf{Sao Paulo} & $0.51 \pm 0.04$ & 0.89 & 0.41 & $0.68 \pm 0.02$ & 0.71 & - & 2.17 \\
        \textbf{Tehran} & $0.55 \pm 0.01$ & 0.10 & 0.54 & $-0.20- \pm 0.01$ & -0.18 & - & 0.19 \\
        \textbf{Las Vegas} & $0.55 \pm 0.02$ & 0.41 & 0.37 & $0.21 \pm 0.01$ & 0.18 & $0.37 \pm 0.07$ & 1.11 \\
        \textbf{Atlanta} & $0.56 \pm 0.02$ & 0.04 & 0.52 & $-0.25 \pm 0.01$ & -0.22 & - & 0.08 \\
        \hline
    \end{tabular}
    \caption{\textbf{Measure of exponents.} $\alpha_{\text{loc}}$ is measured by fitting the first part of the curve $w(l, P) \sim l^{\alpha_{\text{loc}}} $. We average the exponent over the years and give the standard deviation. $\beta$, $1/z$ and $\beta^*$ are measured through the optimal collapse method explained in the Methods. We show measures of $1/z$ with the patches method explained in the Methods. The incertitude corresponds to the 95\% confidence interval. The patches method seem to estimate value slightly lower than the optimal collapse -- however the latter are tendentially close and within the $95\%$ confidence interval on $1/z$ (patches). In the last column, we show the values of $\alpha=z \beta$. }
    \label{tab:exponents}
\end{table}
}

